\documentclass[a4paper,11pt]{article}
\pdfoutput=1 

\usepackage{jheppub} 

\newcommand{\st}{\begin{equation}}
\newcommand{\stp}{\end{equation}}
\def\Eq#1{(\ref{#1})}
\def\Eqs#1{(\ref{#1})}
\def\eq#1{(\ref{#1})}
\def\eqs#1{(\ref{#1})}
\def\app#1{appendix~\ref{#1}}
\def\fig#1{figure~\ref{#1}}
\def\figs#1{figures~\ref{#1}}
\def\Fig#1{Figure~\ref{#1}}
\def\sect#1{section~\ref{#1}}

\def\p{\mathbf{p}}
\def\x{\mathbf{x}}
\def\k{\mathbf{k}}

\newcommand{\be}{\begin{equation}}
\newcommand{\ee}{\end{equation}}

\newcommand{\llangle}{\left\langle}
\newcommand{\rrangle}{\right\rangle}
\def\x{{\mathbf x}}

\newcommand{\ti}{\tau_{\rm init}}
\title{Initial conditions for hydrodynamics from weakly coupled pre-equilibrium 
evolution}
\preprint{CERN-TH-2016-117}
\author[a]{Liam Keegan}
\emailAdd{liam.keegan@cern.ch}
\author[a,b]{Aleksi Kurkela}
\emailAdd{a.k@cern.ch}
\affiliation[a]{Theoretical Physics Department, CERN, Geneva, Switzerland}
\affiliation[b]{Faculty of Science and Technology, University of Stavanger, 
4036 
Stavanger, Norway}
\author[c]{Aleksas Mazeliauskas}
\emailAdd{aleksas.mazeliauskas@stonybrook.edu}
\affiliation[c]{Department of Physics and Astronomy, Stony Brook University, 
Stony 
Brook, New York 11794, USA}
\author[c]{Derek Teaney}
\emailAdd{derek.teaney@stonybrook.edu}
\abstract{
We use effective kinetic theory, accurate at weak coupling, to simulate 
the pre-equilibrium evolution of transverse energy and flow perturbations in 
heavy-ion collisions. We provide 
a Green function which propagates the initial perturbations to the 
energy-momentum tensor at a time
when hydrodynamics becomes applicable. With this map,
the complete pre-thermal evolution from saturated nuclei to hydrodynamics can 
be modelled in a perturbatively controlled way. 
}
\arxivnumber{1605.04287}
\begin{document}

\maketitle
\flushbottom
\section{Introduction}
Viscous relativistic hydrodynamics provides a remarkably detailed and 
phenomenologically successful description of the expansion of the Quark Gluon Plasma (QGP)
in the ultrarelativistic heavy-ion 
collisions realized at the BNL Relativistic Heavy Ion Collider (RHIC) and at 
the 
CERN 
Large 
Hadron Collider (LHC)~\cite{Heinz:2013th,Luzum:2013yya,Teaney:2009qa}.
Hydrodynamics is an effective theory based on an assumption that the medium is
sufficiently close to  local thermal equilibrium
that the full stress tensor can be  
expanded in gradients of the energy and momentum densities~\cite{Baier:2007ix}. 
However, due to
the singular geometry of heavy ion collisions, the gradients diverge at early times, and the hydrodynamic approach does not apply during the initial stages of
the evolution.
Indeed, hydrodynamic simulations start at some sufficiently late
\emph{initialization time} $\ti\sim 1\,\text{fm}/c$, when the gradient 
expansion
becomes a useful approximation scheme. 
The initial conditions for hydrodynamics at $\ti$ are generally unknown, and
must be parametrized and fitted to data~\cite{Bernhard:2015hxa}. This procedure
often neglects any prethermal
evolution, and 
limits the empirical determination of the transport coefficients of the
QGP~\cite{Liu:2015nwa}.

A useful prethermal model  should smoothly and
automatically approach hydrodynamics. If this is the case, 
the combined pre-thermal and hydrodynamic evolutions will be
independent of the initialization
time~\cite{vanderSchee:2013pia,Romatschke:2015gxa,Kurkela:2016vts}. 
In most simulations the prethermal evolution is
either completely neglected \cite{Niemi:2015qia}, or modelled in a way that 
does not contain the
correct physics to produce  hydrodynamic 
flow~\cite{Broniowski:2008qk,Schenke:2012wb,Liu:2015nwa}.
 In addition, in
some models (such as the successful IP-glasma model \cite{Schenke:2012wb} motivated by parton saturation) the initial conditions  
contain strong gradients 
which limit the effectiveness of the hydrodynamic derivative expansion~\cite{Niemi:2014wta,Noronha-Hostler:2015coa}. 
Different hydrodynamic codes regulate these  extreme initial conditions in
different ad hoc ways, e.g. by arbitrarily setting the shear stress
tensor to zero when the hydrodynamics is initialized.  Again, these ambiguities 
limit the ability of hydrodynamic simulations to determine the
transport properties of the QGP.

In the limit of weak coupling $\alpha_s \ll 1$ the approach to hydrodynamics, or \emph{hydrodynamization}, is 
described by an effective kinetic theory (EKT) \cite{Arnold:2002zm}, which 
takes into account the non-trivial 
in-medium dynamics of screening and the Landau-Pomeranchuk-Migdal suppression of
collinear radiation.
In ref.~\cite{Kurkela:2015qoa} (which includes one of the authors), it 
was shown that the EKT, starting with 
initial 
conditions motivated by the Color-Glass Condensate (CGC) saturation framework~\cite{Iancu:2002xk,Iancu:2003xm,Gelis:2010nm, Gelis:2007kn, Lappi:2011ju}, reaches hydrodynamics in a phenomenologically reasonable time
scale of $\sim 10/Q_s$, where $Q_s$ is the (adjoint representation) saturation scale, which is estimated to be of order of few GeV
for central heavy-ion collisions at the LHC.  This first calculation used the EKT
to monitor the equilibration of a uniform plasma of infinite transverse extent  during a Bjorken expansion.

Transverse gradients in the profile 
will initiate flow during the equilibration process. This \emph{preflow}
and the accompanying modifications
of  the initial energy density profile will influence the subsequent hydrodynamic evolution. The goal of the current paper is to use the EKT 
to precisely determine the preflow and the components of the 
energy momentum tensor $T_{\mu}^{\phantom{\mu}\nu}(\ti, {\bf x}_\perp)$ 
that should be used to initiate the hydrodynamic evolution for
a given initial energy density profile.
Although the kinetic theory calculation can be used to match different models for the initial energy profile to hydrodynamics,  the weak coupling
approximations made in the IP-glasma model lead naturally to
effective kinetic theory.

\begin{figure}
    \begin{center}
        \includegraphics[width=0.7\textwidth]{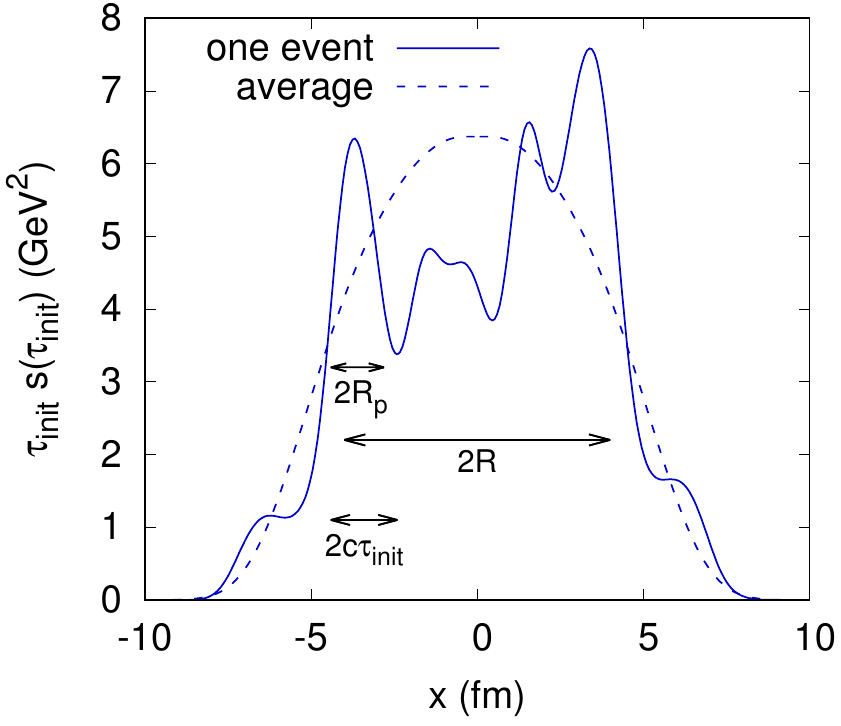}
        \caption{A typical entropy density profile (times ${\ti}$) for 
            a single event used 
            as an initial condition in 
            current hydrodynamic simulations at the LHC for a 0-5\% centrality class~\cite{Mazeliauskas:2015vea}.  
            An
        event averaged initial condition is shown by the dashed line. 
            Often the initial flow velocity is set to zero. The different 
            scales are discussed in the text.
    \label{ichydro}}
    \end{center}
\end{figure}

\Fig{ichydro} shows a typical transverse (entropy) profile that is used  in
current hydrodynamic simulations~\cite{Mazeliauskas:2015vea}. Clearly
during the equilibration process the profile will change and generate initial 
flow.
The equilibration time, $c\ti$, is short compared to the nuclear radius,  $R$.
For this reason the prethermal evolution is
insensitive to the global collision geometry.
Indeed, we may decompose the transverse plane into
causally disconnected patches of size $c\ti \ll R $  whose prethermal evolution 
can be separately determined. 
In these patches, the global nuclear geometry determines
a small gradient that can be considered as a linear perturbation over a translationally invariant background. 
Thus, corrections to initial conditions for hydrodynamics
from the global geometry are of order $c\ti/R$~\cite{Vredevoogd:2008id}.
In addition to the global geometry, the initial energy density profile includes event-by-event fluctuations at smaller 
scales set by the nucleon size $R_p$,  which is comparable to the 
causal horizon $R_p \sim c \ti$.
Event-by-event fluctuations at these length scales are  suppressed by 
$1/\sqrt{N_\text{part}}$ where $N_{\text{part}}$ is the number of participating 
nucleons in the event,  $N_{\text{part}}\sim 100-300$.
Therefore,  such fluctuations can also be treated
in a linearized way as fluctuations over a translationally invariant
background.
The
structure of the initial profile at even smaller
scales is less well known, but in models based on CGC, one expects fluctuations 
to subnuclear scales of order the saturation momentum, $Q_s^{-1} \sim 
0.1\,\text{fm}$. 

Finally, 
an important scale is set by the mean free path, which in a weakly 
coupled theory is of order $1/\lambda^2 T_{\rm eff}$ for states
not too far from equilibrium.  In practice, this length scale is comparable,
though slightly shorter than the causal horizon and the nucleon scales. Without the scale separation, the medium prethermal response  to initial perturbations in the transverse plane can only be 
computed by a calculation within the EKT. Fortunately, as discussed
above linearized kinetic theory is sufficient to determine this response. 

To summarize, our strategy is to use linearized kinetic theory to follow 
the hydrodynamization  of perturbations on top of a far-from-equilibrium  Bjorken background with translational symmetry in the transverse directions.  This determines
the stress tensor for hydrodynamics  at the initialization time.
The length scales of relevance are 
the nuclear-geometry, the nucleonic scale, the causal horizon $c\ti$, and the mean free path
\begin{equation}
    R \gg R_{p} \sim c\ti \sim \frac{1}{\lambda^2 T_{\rm eff}} \, .
\end{equation}
By linearizing the problem and solving for the response,
we will determine a Green function describing how an energy fluctuation  
at the earliest moments, $\tau\sim 1/Q_s$,
evolves  during the equilibration process to the  hydrodynamic
fields, i.e. the energy and momentum 
densities, $\delta T^{00}(\ti, \x_\perp)$ 
and $\delta T^{0i}(\ti,\x_\perp)$ respectively. We will verify 
that  the subsequent evolution is described by second order hydrodynamics to
certifiable precision.

In \sect{linearized} we outline the linearized EKT, and study the linear 
response of the EKT in equilibrium. In \sect{Bjorken} we systematically study
the approach to equilibrium of Fourier modes of specified $k$, starting
with a far from equilibrium initial state.  
In \sect{Green} we Fourier 
transform these results and determine a coordinate space Green function which
produces the appropriate
initial conditions for hydrodynamics at $\ti$
when convolved with a specified initial state. 
We also analyze
the long wavelength limit of these Green functions, making contact and providing additional insight into previous work on preflow~\cite{Vredevoogd:2008id}. Finally,
we discuss our conclusions in \sect{discuss}.

\section{Linearized kinetic theory}
\label{linearized}

\subsection{Setup}
At weak coupling the non-equilibrium evolution of the boost invariant color and 
spin averaged
gluon distribution function
is described in 
terms of an effective kinetic equation \cite{Arnold:2002zm}
\begin{align}
\partial_\tau f_{\x_\perp,\p}+ \frac{\bf p}{|p|}\cdot \nabla_{\x_\perp} 
f_{\x_\perp,\p} - \frac{p_z}{\tau}\partial_{p_z} f_{\x_\perp,\p}= 
-\mathcal{C}[f_{\x_\perp,\p}],
\end{align}
where the effective collision kernel $\mathcal{C}[f]$ incorporates the elastic $2\leftrightarrow 2$ and inelastic $1\leftrightarrow 2$ processes as required for a leading order description in the coupling constant $\lambda =4\pi \alpha_s N_c$, which is the only parameter of the EKT. 
The kinetic theory is valid when the occupancies are perturbative $\lambda f \ll 1$ and when the relevant
distance scales are larger than the typical Compton wavelength of the particles 
$\Delta x \gtrsim \langle p \rangle^{-1}$. 
The details 
of the scattering kernel have been discussed in refs.~\cite{Arnold:2002zm, 
Kurkela:2015qoa,Keegan:2015avk} and are briefly repeated here in the 
\app{colker}.
We use the isotropic screening approximation from \cite{Kurkela:2015qoa} which is leading order 
accurate  for parametrically isotropic systems $\mathcal P_L/\mathcal P_T \approx 1$. (Here and below $\mathcal P_{L}$ and $\mathcal P_T$ denote the longitudinal and transverse pressures.)
In the current paper we will consider only gluonic degrees of freedom and assume that 
the contribution of quarks is suppressed during the pre-equilibrium evolution.%
\footnote{The initial far-from-equilibrium state is parametrically dominated by gluons. Once the plasma
has thermalized it should contain also fermionic degrees of freedom. However, the production of fermions is suppressed
by larger color factors $C_F/C_A$,  and 
by Pauli blocking factors (while scattering of gluons is Bose enchanced). 
It is therefore plausible that the system hydrodynamizes before
it is chemically equilibrated.}

We split the 
distribution function into a translationally symmetric background and a 
linearized perturbation with a 
wavenumber $\k_\perp$ in the transverse plane
\begin{equation}
f_{\x_\perp,\p}=\bar f_\p + \int \frac{d^2 \k_\perp}{(2\pi)^2} A({\bf k}_\perp) 
e^{i \k_\perp \cdot \x_\perp}\delta f_{\k_\perp,\p},
\end{equation}
where $A({\bf k}_\perp)$ characterizes the initial density profile. The kinetic 
equations for the background
and the (complex) fluctuation then read
\begin{subequations}
\label{boltz}
\begin{eqnarray}
(\partial_\tau - \frac{p_z}{\tau} \partial_{p_z})  \bar f_{\p} &=& - \mathcal{C}[\bar f], \\
(\partial_\tau - \frac{p_z}{\tau} \partial_{p_z} + \frac{i \p_\perp\cdot \k_\perp}{p} )  \delta f_{\k_\perp,\p} &=& - \mathcal{C}[\bar f, \delta f],
\end{eqnarray}
\end{subequations}
where $\mathcal{C}[\bar f, \delta f]$ is the collision kernel linearized in 
$\delta f$ (see \app{colker}  for details).

\begin{figure}
\centering
\includegraphics[width=0.7\textwidth]{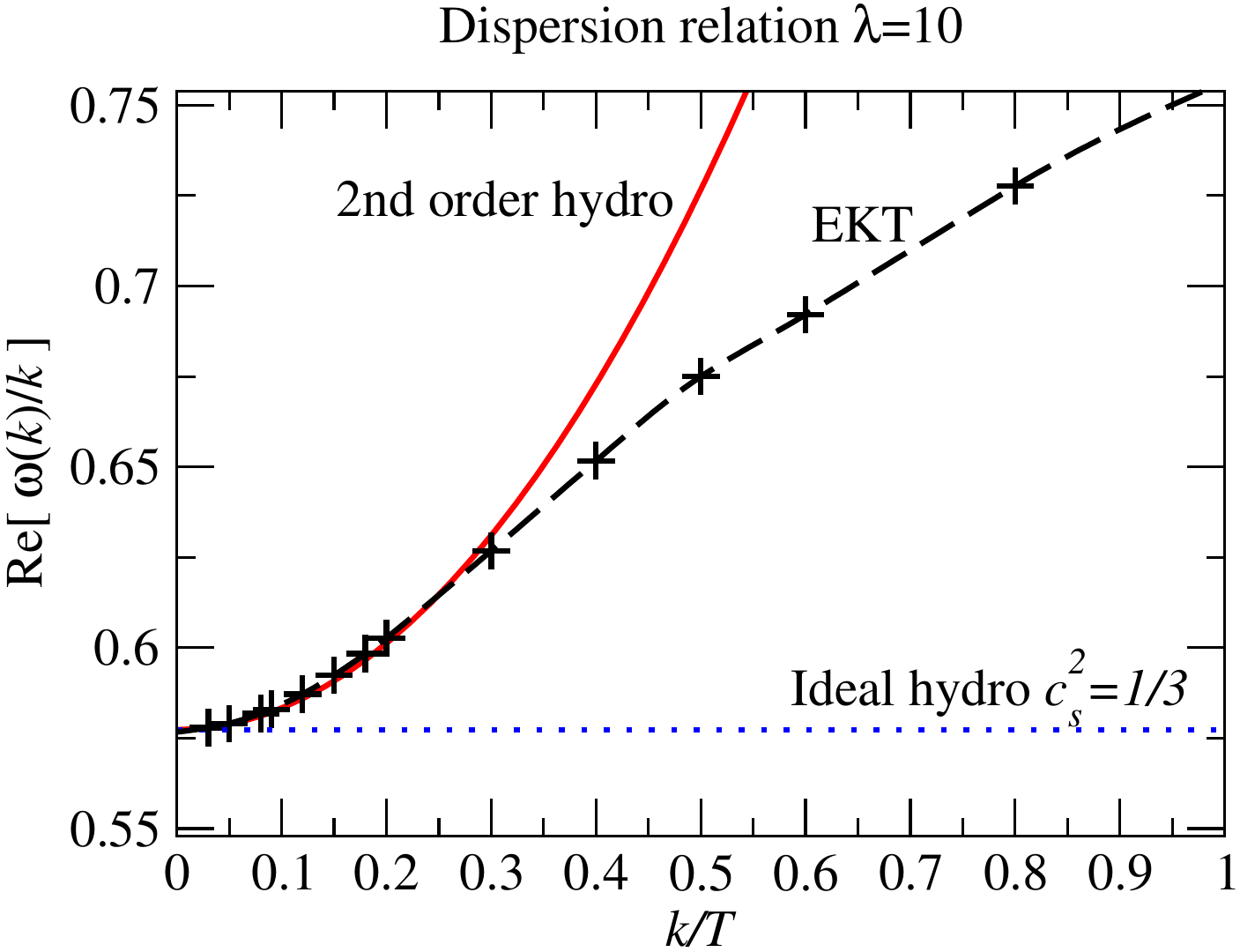}
\caption{\label{disprel} The dispersion relation of sound modes with thermal background from the EKT.
The long wavelength modes are described by ideal hydrodynamics with $\omega = c_s k$ and $c_s^2 = 1/3$, and the approach to ideal 
hydrodynamics is well described by 2nd order hydrodynamics. For modes with wave numbers larger than $k \gtrsim 0.4 T$,
the dispersion relation differs significantly from the hydrodynamic expectation.
}
\end{figure}

\subsection{Hydrodynamization close to equilibrium}
\label{kspace}
Before studying the equilibration  process,
we will analyze the linear response of the EKT close
to equilibrium, corresponding to the $\tau\rightarrow \infty$ limit of \eq{boltz}.
Our goal in this section is to determine at what wavenumbers (characterized by $k/T$) linearized energy-momentum perturbations  are described by hydrodynamics for an equilibrated background.

The 
dispersion relation for the sound mode to second order in the hydrodynamic expansion 
reads~\cite{Baier:2007ix}
\begin{equation}
\omega = c_s k - i\frac{4}{3}\frac{\eta }{e+p}k^2 + \frac{4}{3}\frac{\eta}{e+p}\left( c_s \tau_\pi - \frac{2}{3 c_s}\frac{\eta}{e+p}\right)k^3,
\label{eq:disp_rel}
\end{equation}
where $c_s^2 = 1/3$ for conformal equation of state and $\eta, \tau_\pi$, are known transport coefficients at weak coupling \cite{Arnold:2003zc, York:2008rr}. For $\lambda = 10$ (corresponding to $\alpha_s \approx 0.26$) the hydrodynamic coefficients read $\eta/s =  0.62$, 
$\tau_\pi = 5.1 \eta/sT$, and $\lambda_1 = 0.8 \eta \tau_\pi$.
We will quantify at what numerical 
values of  $k/T$ the corrections to \eq{eq:disp_rel} become sizeable. 

To this end, the kinetic theory is initiated  in local 
thermal equilibrium with a spatially varying temperature, 
$T(\x) = T + \delta T e^{i \k_\perp \cdot \x_\perp}$, and  corresponding  phase space distribution
\begin{equation}
\delta f^{(1)}_{\k_\perp,\p} =  -\frac{\delta T}{T} p \, \partial_p \bar 
f_{\p}, \quad \textrm{ and}\quad
\bar f_\p = \frac{1}{e^{p/T}-1}.
\end{equation}
Gradients in the energy density drive
momentum perturbations in due course, and the
frequency of the subsequent (damped) oscillations determines the real part of the
dispersion relation. Numerically, we obtained the oscillation frequency by measuring the time
interval between the successive nodes. 

The results
for various $k/T$ are depicted in \fig{disprel} for $\lambda =10$.  We see that 
at small $k$,  $k/T \lesssim 0.1$, the 
dispersion relation is well described by the $c_s^2 = 1/3$ result of ideal hydrodynamics, and the approach to ideal hydrodynamics
is described by the 2nd order corrections of \eq{eq:disp_rel}
(note that 
the real part of the frequency does not get a first order correction). 
Indeed, 
for $k/T \lesssim 0.4$, the second order hydrodynamic theory matches well 
with the EKT. 

At higher values of $k/T$, the EKT finally saturates at $\omega = 
k $ in contrast to the strict (unresummed) second order hydrodynamics. For $\lambda = 10$ this 
happens 
only 
at rather large values of $k$, $k > T$. For these wavenumbers, the wavelength of the perturbation is  comparable  to the typical gluon 
Compton wavelength, and the linear response of the system
cannot be reliably computed with kinetic theory in this regime.

We conclude that for $\lambda = 10$, the smallest scales that hydrodynamize have $k \sim 0.4 \, T$. Varying the 
value of $\lambda < 10$ (not shown), we find that the scale where hydrodynamics breaks downs tracks the shear viscosity, $k \sim 0.4 \, [\eta(\lambda = 10)/\eta(\lambda)]\,T $ with varying $\lambda$. Note that for smaller $\lambda$, the saturation to $\omega \sim k$ can take place within the
regime of validity of the effective theory.

\section{Hydrodynamization of fluctuations far from equilibrium}
\label{Bjorken}
We now move on to study the hydrodynamization of spatially dependent fluctuations 
on top of a far-from-equilibrium boost invariant background.
As discussed in \cite{Kovner:1995ja, Lappi:2006fp, Gelis:2013rba, Kurkela:2015qoa}, at very 
early times $\tau \lesssim Q_s^{-1}\,$ the 
far-from-equilibrium gluonic system in the midrapidity region is parametrically over-occupied $\lambda f \sim 1$,
and the dynamics is described with coherent classical gauge fields rather than with particles. This part of the evolution is 
characterized by negative values of the longitudinal pressure $\mathcal P_L$, which is a result of the coherence of
the approximately boost invariant fields. However, classical numerical
simulations~\cite{Lappi:2011ju, Gelis:2013rba} (as well as analytical series
solutions to the classical equations of motion~\cite{Chen:2015wia,Li:2016eqr})
show that  in a timescale $Q_s \tau \sim 1$, 
the coherence is lost, the longitudinal pressure approaches zero $\mathcal P_L \sim 0$, and the occupancies become perturbative \cite{Kurkela:2011ti, Kurkela:2011ub,Berges:2013fga, Baier:2000sb}. At this point the system may be passed
to the EKT \cite{Mueller:2002gd,Jeon:2004dh, York:2014wja, Kurkela:2015qoa}. 

Following \cite{Kurkela:2015qoa}, we take as our initial condition at $\tau_0 = 1/Q_s$ a parametrization
\begin{align}
    f(p_z,p_\perp) &= \frac{2}{\lambda} A f_0(p_z \xi/p_0,p_\perp/p_0), \label{eq:init_cond}\\
f_0(\hat p_z,\hat p_\perp) & = \frac{1}{\sqrt{\hat p_\perp^2+\hat p_z^2}} e^{-2 (\hat  p_\perp^2 + \hat  p_z^2)/3}, \label{f0}
\end{align}
where $p_0 = 1.8\,Q_s$, $\xi=10$, and $\lambda=10$.  
The parameters $p_0$ and $\xi$ are motivated 
by classical simulations where $\sqrt{\llangle p_T^2 \rrangle}\approx 1.8\, Q_s$ 
and $\llangle p_z^2 \rrangle \ll \llangle p_T^2 \rrangle$.
The amplitude, $A$, is adjusted so 
that energy per rapidity
\begin{align}
    \tau_0 e(\tau_0) =  \tau_0 \nu_g \int \frac{d^3p}{(2\pi)^3} \, |p|\, f(p_z, p_\perp) \, ,
\end{align}
matches the results
of classical simulations \cite{Lappi:2011ju}, where
\begin{align}
    \label{Qsdef}
    \tau_0 e(\tau_0)  \simeq  0.358\,\frac{\tau_0 \nu_g Q_s^4}{\lambda} \, .
\end{align}
Here $\nu_g = 2 d_A= 16$ is the number of gluonic degrees of freedom.
With these parameters, the number of gluons and the mean $p_T$ in the EKT at $\tau_0$  are
\begin{align}
    \frac{dN}{d^2\x_\perp dy} =&  0.232\,\frac{\nu_g Q_s^2}{\lambda} ,  \qquad \sqrt{\llangle p_{T}^2\rrangle}  = 1.8\,Q_s,
 \end{align}
 which roughly matches the classical Yang-Mills simulations.

We will follow 
the response to two specific initial perturbations of these initial conditions,
which provide an independent basis for describing   energy and momentum
fluctuations in the transverse plane. For the energy fluctuation, we take
\begin{align}
\delta f^{(1)}_{\k_{\perp},\p}(\tau_0) &= - \frac{\delta Q_s}{Q_s} p \, 
\partial_p \bar 
f_{\p} \label{initialdf},
\end{align}
which results from varying 
the saturation scale in \eq{eq:init_cond}, $Q_s(\x_\perp) \sim Q_s + \delta Q_s e^{i 
\k_\perp \cdot \x_\perp}$. This initial condition is studied
    in the bulk of the paper. For the momentum fluctuation
    we take
\begin{align}
\delta f^{(2)}_{\k_{\perp},\p}(\tau_0) &= -\frac{\delta Q_s}{Q_s} \, 
\hat{\k}_\perp\cdot 
\partial_{\bf p} \bar f_{\p},\label{initialdf2}
\end{align}
which does not perturb the local energy density.
The response to this initial condition is recorded in \app{velpert}.

\begin{figure*}
\centering
    \includegraphics[width=1\textwidth]{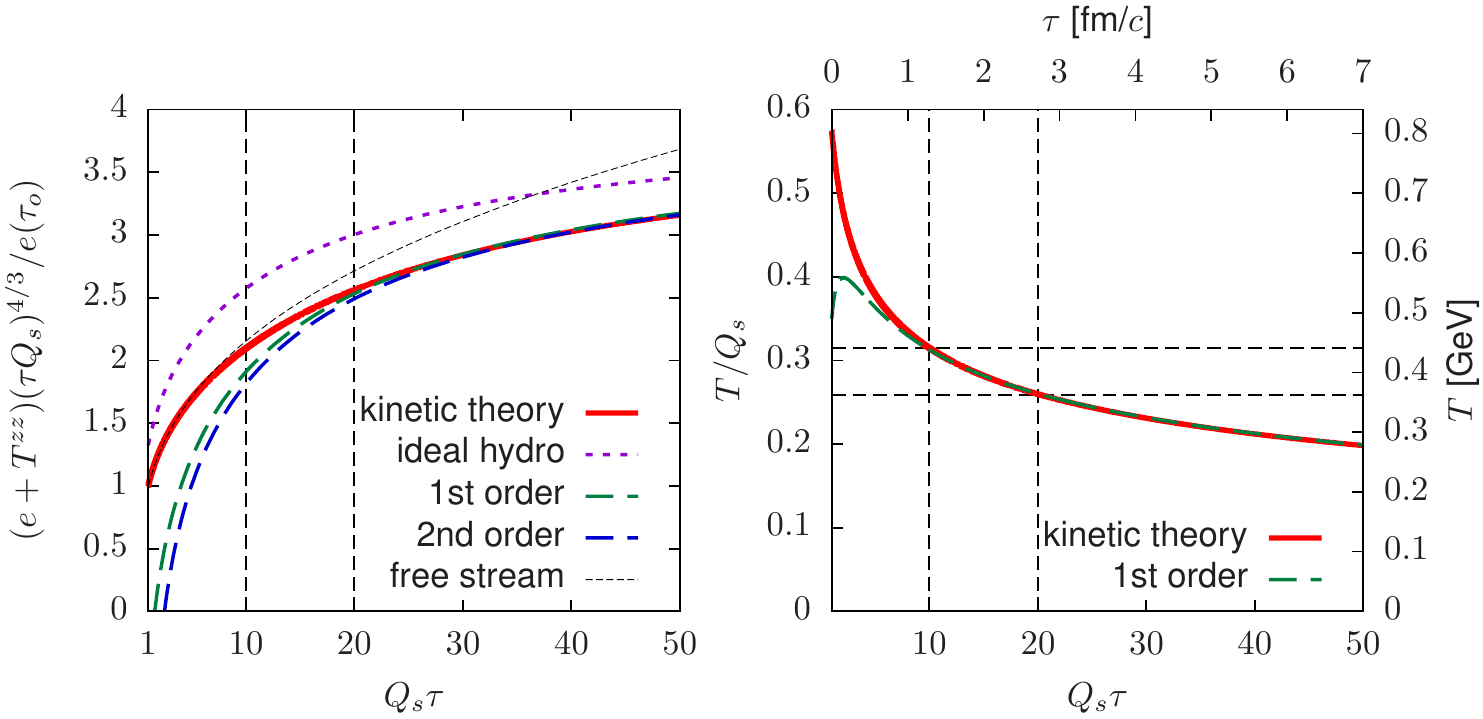}
\caption{(a) A comparison of the relevant combination $e(\tau) + T^{zz}(\tau)$ for the  
kinetic theory background with the hydrodynamic constitutive equations of \eq{eqTzz}.
(b) The background effective temperature as obtained from the Landau matching condition
$e = \nu_g \frac{\pi^2}{30} T^4$.  Extrapolation of first order hydro (fitted 
at asymptotic late 
times) is shown for comparison, \Eq{asympT}. The scales in physical units 
correspond to 
$Q_s = 1.4\,{\rm GeV}$ which yields the entropy required by hydrodynamic 
simulations (see text).
    \label{background}
}
\end{figure*}

Without loss of generality, we can choose the wave vector $\hat{k}_\perp=(k,0)$ 
to 
point in $x$-direction. Then at any time, the energy and momentum 
perturbations are defined 
as
\begin{align}
\delta e(\tau,k)  & \equiv \delta T^{00}=\nu_g \int \frac{d^3\p}{(2\pi)^3 
}p^0 \delta f,\\
g^x(\tau, k) &\equiv \delta T^{0x} =  \nu_g \int \frac{d^3\p}{(2\pi) 
^3}p^x \delta f,
\end{align}
and their evolution is governed by the linearized conservation equations
\begin{subequations}
    \label{eq:conservation}
\begin{align}
&\partial_\tau e(\tau)=-\frac{e(\tau)+
T^{zz}(\tau)}{\tau}, \label{econs}\\
&\partial_\tau \delta e(\tau,k)+ik  g^x (\tau,k)=-\frac{\delta 
e(\tau,k) +  \delta T^{zz}(\tau,k)}{\tau},\label{decons}\\
&\partial_\tau g^x(\tau,k)+ik \delta T^{xx}(\tau,k)= -\frac{g^x 
(\tau,k)}{\tau},\label{dgcons}
\end{align}
\end{subequations}
where $\delta T^{\mu\nu}$ is the energy-momentum tensor perturbation caused by 
$\delta f$.  If the system is described by  
hydrodynamics then $T^{zz}$, $\delta T^{zz}$ and $\delta T^{xx}$ 
are determined through the constitutive equations by the first moments of the particle distribution function $e$, 
$\delta e$ 
and $g^x$. For conformal second order 
viscous hydrodynamics these relations are
\begin{subequations}
    \label{eq:constit}
\begin{align}
T^{zz}(\tau) &= 
\frac{1}{3}e-\frac{4}{3}\frac{\eta}{\tau}-\frac{8}{9}\frac{\tau_\pi 
\eta - 
\lambda_1}{\tau^2},\label{eqTzz}\\
\delta T^{xx}(\tau,k) = &\frac{\delta e(\tau, k)}{e} 
\left[\frac{1}{3}e + 
\frac{1}{3}\eta \tau_\pi 
k^2 + \frac{1}{2\tau}\eta -\frac{2\left( 
\lambda_1-\eta\tau_\pi\right)}{9\tau^2} \right] \nonumber \\
&- i \frac{kg^x(\tau,k)}{e}\left[
\eta-\frac{1}{\tau}\left( \frac{\eta^2}{2e} + \frac{\eta 
\tau_\pi}{2} -\frac{2}{3}\lambda_1 \right)\right]\label{eqdTxx}\\
\delta T^{zz}(\tau,k) = &\frac{\delta e(\tau, k)}{e} 
\left[\frac{1}{3}e - 
\frac{1}{6}\eta \tau_\pi 
k^2 - \frac{1}{\tau}\eta +\frac{4\left( 
\lambda_1-\eta\tau_\pi\right)}{9\tau^2} \right] \nonumber \\
&+ i \frac{kg^x(\tau,k)}{e}\left[
\frac{1}{2}\eta-\frac{1}{\tau}\left( \frac{\eta^2}{4e} +\frac{2}{3}\lambda_1 
\right)\right]\label{eqdTzz},
\end{align}
\end{subequations}
where  the constitutive equations for ideal (or first order viscous) hydrodynamics 
can 
be recovered by 
setting $\eta=\tau_\pi=\lambda_1=0$  (or $\tau_\pi=\lambda_1=0$).

\subsection{Evolution of the background energy density}

Before studying the perturbations, we will study the equilibration of
the background energy density, elaborating on  the original 
study~\cite{Kurkela:2015qoa}. In \fig{background}(a) we compare the energy 
momentum tensor combination $e+T^{zz}$ 
in  the kinetic theory simulation
to the constitutive equation, \eq{eqTzz}. The $e+T^{zz}$ combination is motivated by the  conservation law in \eq{econs}. 

At early times the system evolves approximately according to  free streaming%
\footnote{During this part of the evolution, the system evolves according to the nonthermal attractors
discussed in e.g. \cite{Baier:2000sb,Kurkela:2011ub,Berges:2013fga,Kurkela:2015qoa}. The nonthermal attractors are characterized by  $T^{zz}\ll e$, and for the current discussion the fine details of the attractor are irrelevant, and the evolution resembles that of free streaming.},
with $T^{zz}\sim 0$ and $e \propto \tau^{-1}$.
As already noticed in \cite{Kurkela:2015qoa}, the constitutive equations give 
an increasingly accurate description of the EKT
stress tensor as a function of time. While the ideal constitutive equations are 
rather far from the EKT at all relevant times, the viscous and 2nd order 
equations quickly converge to the EKT. Note that an accidental (approximate)
cancellation of $\lambda_1 - \eta \tau_\pi$ makes the 
second order correction anomalously small~\cite{Luzum:2008cw}, and only 
at rather late times  after non-hydrodynamic modes have
almost completely decayed does second order hydrodynamics finally improve the first order result (not shown).
By times $Q_s
\tau =\{ 10, 20 \}$, the viscous constitutive equations agree
with the EKT within $\{10\%,2\%\}$.

It is noteworthy that the EKT interpolates smoothly between the free streaming
and viscous hydrodynamic evolutions without an extended period during which the
evolution is not approximately described by one or the other approximation
scheme. At $Q_s\tau{=}10$ the evolution is somewhat closer to free streaming,
and using hydrodynamics at this point is a rough, though perhaps acceptable,
approximation. At $Q_s\tau{=}20$ hydrodynamics is a better approximation, but at this time the causal horizon is becoming comparable to the nuclear radius.

Given the agreement with the constitutive equations, one  can use
the hydrodynamic equations to propagate the system forward in time. 
At late times ideal hydrodynamics is valid, and the entropy per area 
per rapidity approaches a constant
\st
\label{finalentropy}
\lim_{\tau \rightarrow \infty} \tau s(\tau) \equiv \frac{\nu_g \Lambda_s^2}{\lambda}.
\stp
(Here the $\Lambda_s$ parametrization is motivated by the scaling of the 
initial multiplicity with the saturation scale in \eq{Qsdef}.)
Dimensional reasoning indicates that $\Lambda_s^2$ is proportional to 
$Q_s^2$. Taking the data presented in \fig{background}(a) we may extrapolate
  $\tau \rightarrow \infty$ to determine the proportionality coefficient 
\st
\label{asymptotic}
\Lambda_{s}^2 = 1.95 \, Q_s^2.
\stp
Here we have used the ideal equation of state to convert energy density 
to entropy density.  Since the entropy per gluon of an ideal gluon 
gas is 
$
3.6$, \eq{asymptotic} implies the asymptotic number of gluons per area  per rapidity
is more than a factor of two larger than the input number of gluons at $\tau_0$
\st
\left. \frac{dN}{d^2\x_\perp d\eta}  \right|_{\tau\rightarrow\infty}  
=2.33\left. \frac{dN}{d^2\x_\perp d\eta}  \right|_{\tau=\tau_0} .
\stp
At $Q_s \tau = \{10, 20\}$ the entropy and gluon multiplicity have reached only \{72, 82\}\% of their asymptotic values, corresponding to  gluon multiplication factors of $\{1.6,1.9\}$ respectively.  

Finally, let us make phenomenological contact with more complete hydrodynamic simulations of
heavy ion collisions, and estimate the saturation momentum required by 
phenomenology.
The initial entropy in hydrodynamics is normally adjusted to 
reproduce the mean multiplicity. Using the computer code from one such 
hydrodynamic simulation at the LHC~\cite{Mazeliauskas:2015vea},  we computed 
the average  entropy per area at the hydrodynamic initialization 
time\footnote{Specifically, for the event-by-event 
hydro code described in ref.~\cite{Mazeliauskas:2015vea} we first 
    created a smooth entropy density profile, $\overline {s(\x_\perp)} $, by
    averaging over events for a 0-5\% centrality class, $b=[0,3.3]\,{\rm fm}$. 
    This average is shown in \fig{ichydro}. We then computed a single averaged 
    entropy
density by averaging $\overline{ s(\x_\perp) }$ with 
$\overline{s(\x_\perp)}$ as a radial weight. } 
\st
\llangle \ti s (\ti)  \rrangle  = 4.13\,{\rm GeV}^2.
\stp
Entropy production during the subsequent hydrodynamic evolution
is small, approximately 15\%, and therefore this constant is approximately 
independent of the  initialization time. With \eqs{finalentropy} and 
(\ref{asymptotic}),
setting $Q_s\simeq 1.4\,{\rm GeV}$ in the EKT will roughly reproduce the entropy
in hydrodynamic simulations provided the system is passed to hydrodynamics at 
$Q_s \tau=10$  where the entropy in the EKT has reached 70\% of its asymptotic 
value in \eq{finalentropy}. 

In \fig{background}(b) we present the time evolution of the effective 
temperature (determined from the energy density and the equation of state)
in physical units for $Q_s \simeq 1.4\,{\rm GeV}$. Its time dependence at late 
times is well described by first order viscous hydrodynamics with the 
asymptotic value
\begin{equation}
\label{asympT}
\lim_{\tau \rightarrow 
\infty}\left(T+\frac{2}{3}\frac{\eta}{s\tau}\right)\tau^{1/3}= 0.763\, 
Q_s^{2/3}.
\end{equation}
The temperature at $Q_s\tau{=}10$ is $T\simeq 430\,{\rm MeV}$,
or close to three times the pseudo-critical temperature. At these
temperatures, modern weakly coupled techniques can be expected to
work reasonably, justifying our approximation scheme. Similarly,
 with this value of $Q_s$
an initialization time of $Q_s\tau\simeq 10$  corresponds to $\ti \simeq 
1.4\,{\rm fm}$.  
The elliptic and 
triangular flows in central events develop on a later time scale, of order 
$R/c_s \sim 8\,{\rm fm}$, and therefore
an initialization time of this order may be acceptable.
 A more complete study 
including fermions will be needed to definitively answer this question. 

\subsection{Evolution of the perturbations}
\begin{figure*}
    \begin{center}
\includegraphics[width=0.47\textwidth]{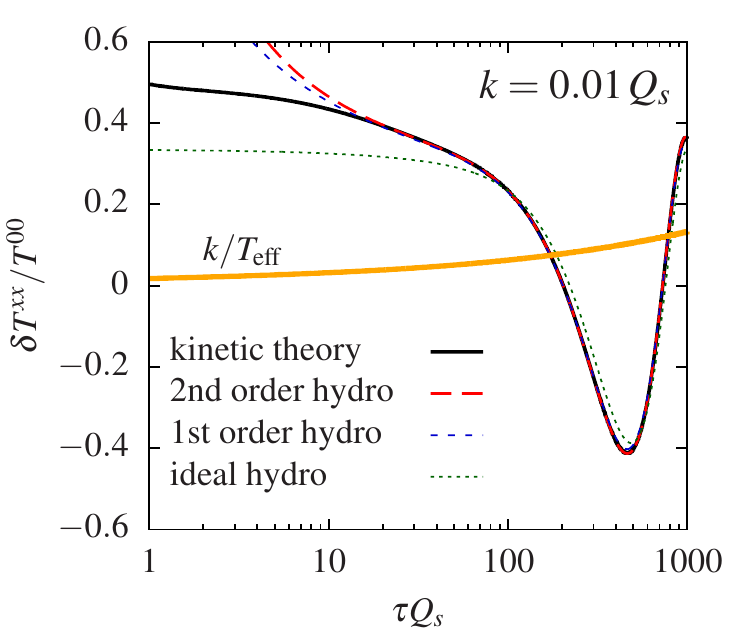}
\includegraphics[width=0.47\textwidth]{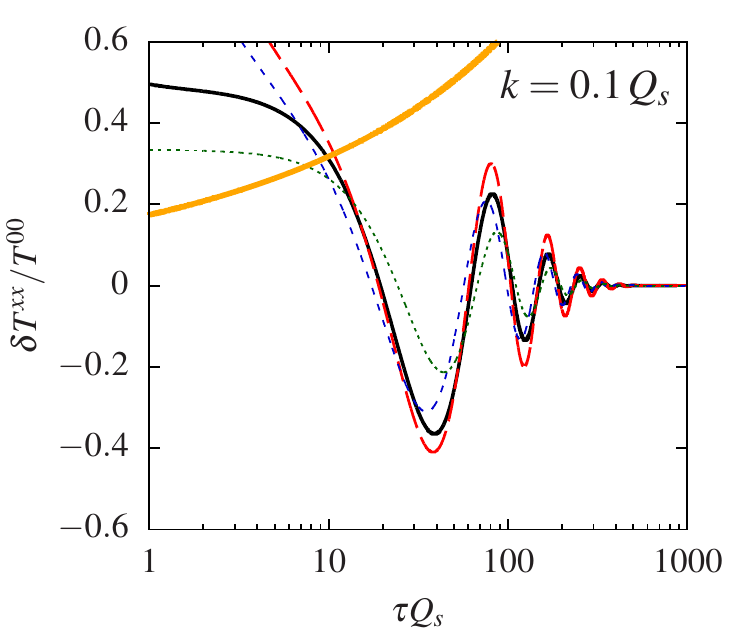}\\
\includegraphics[width=0.47\textwidth]{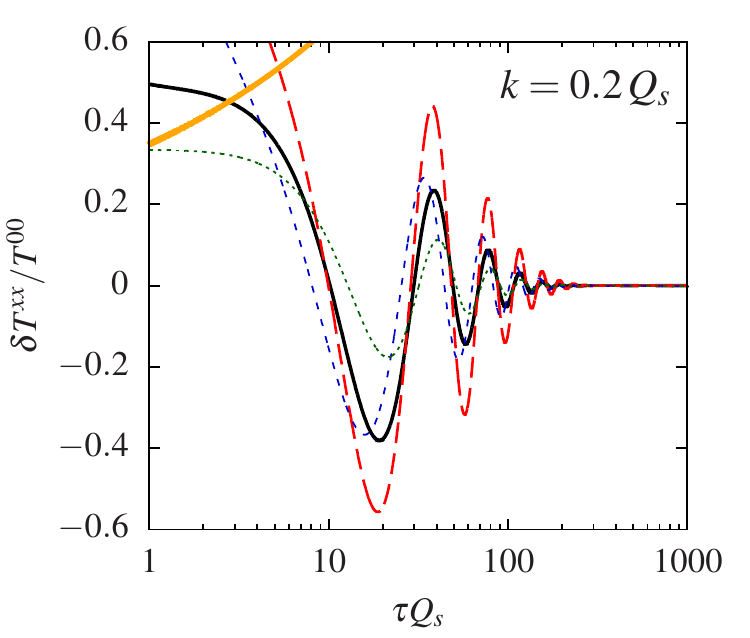}
\includegraphics[width=0.47\textwidth]{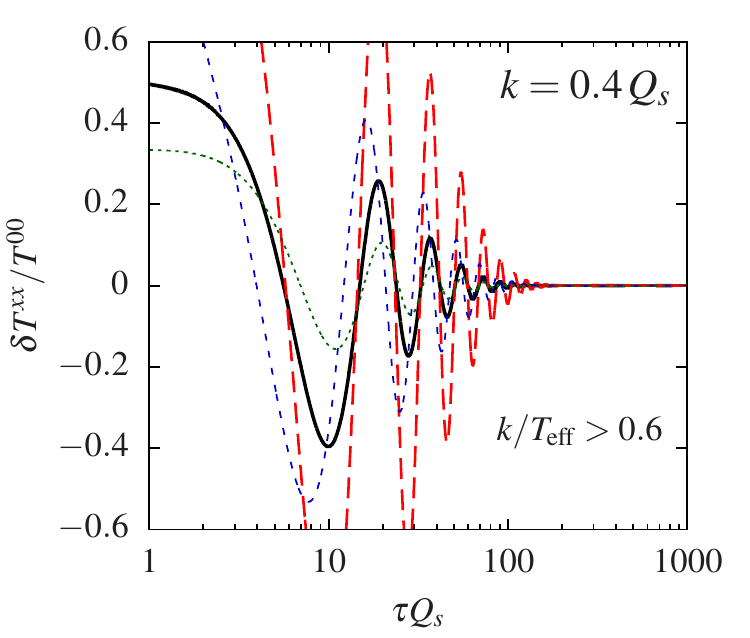}
\end{center}
\caption{$\delta T^{xx}/ T^{00}$ compared with hydrodynamic constitutive
equations (the curves have been normalized by the magnitude of
the initial perturbation $\delta T^{00}(\tau_0)/T^{00}(\tau_0)$). Long wavelengths with  $k\lesssim 0.1\,T$ are
described by the hydrodynamics at approximately the same time as the background
$Q_s \tau \sim 10$. Shorter wavelengths with $k\sim 0.4\,Q_s$ are never well
described by hydrodynamics.  }\label{figdTxx}
\end{figure*}

We now move on to describe the evolution of the linearized perturbations on top
of the thermalizing non-equilibrium background. We first follow the evolution 
of an energy density perturbation, which we start with the initial condition of \eq{initialdf} with different values of $k/Q_s$.
In figures~\ref{figdTxx}(a)-(d) we show $xx$\nobreakdash-component of 
perturbation energy 
momentum tensor $\delta T^{xx}$ compared with ideal, viscous, and second order 
hydrodynamic constitutive equations of \Eq{eqdTxx}.
The lines have been normalized by the background energy density $T^{00}(\tau)$, so that any observed damping 
is due to nontrivial dynamics associated to the spatial inhomogeneity. 

We consider fixed values of $k/Q_s$, which do not correspond to
fixed values of $k/T$, as the effective temperature is changing due to 
the expansion (see \fig{figdTxx}). 
For  very large wavelengths with $k/Q_s = 0.01$, we  
observe  that while ideal constitutive equations have rather 
large corrections, these are well accounted for by the viscous and 2nd order 
equations, roughly at the same time as the background constitutive equation is 
satisfied, i.e. at times after $\tau \sim 10/Q_s$. Determining the temperature 
of 
the background from Landau matching condition $T^4 = \nu_g\frac{\pi^2 }{30} e$ 
(see \fig{background} (right)), we find that at times $\tau = \{10, 20\}/Q_s$, 
the 
wavelength $k=0.01Q_s$ in units of temperature $T=\{  0.31,0.26 \}Q_s$ is 
$k/T(\tau) =  \{0.032, 0.039 \}$. 
As discussed in \sect{kspace} and in \fig{disprel} these values 
of $k/T$ are accurately described by second order hydrodynamics.

We see that even for larger $k/Q_s = \{0.1,0.2,0.4\}$ the hydrodynamic constitutive relations are approximately 
fulfilled when the background has hydrodynamized around $\tau \sim 10/Q_s$. However, larger values of $k/Q$ 
correspond to larger values of $k/T$, and even at late times there are corrections to the constitutive equations. While these 
corrections are moderate for $k/Q_s=0.2$ for which $k/T(\tau = 10/Q_s) \approx 0.6$, they remain $\mathcal{O}(1)$ for $k/Q_s = 0.4$.
It is therefore questionable whether it is justifiable to pass these short scales to hydrodynamic description at any time. 

\section{A Green function for hydrodynamics}
\label{Green}

We now move on to describe how the response to the linearized perturbations in 
EKT can 
be used in a hydrodynamic simulations to encapsulate the far-from-equilibrium 
dynamics of transverse perturbations
during the time scales between $\tau \sim 1/Q_s$ and $\{10,20\}/Q_s$.

\begin{figure*}
\includegraphics{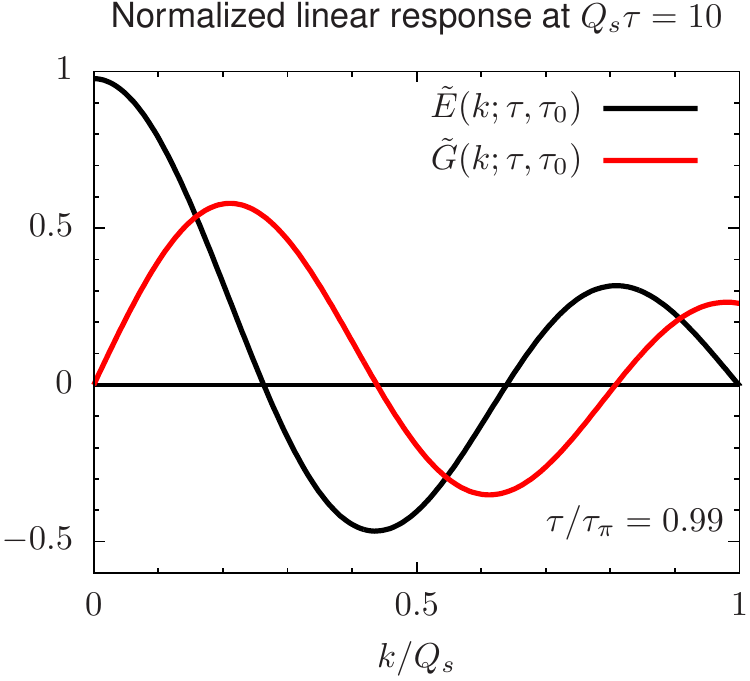}
\includegraphics{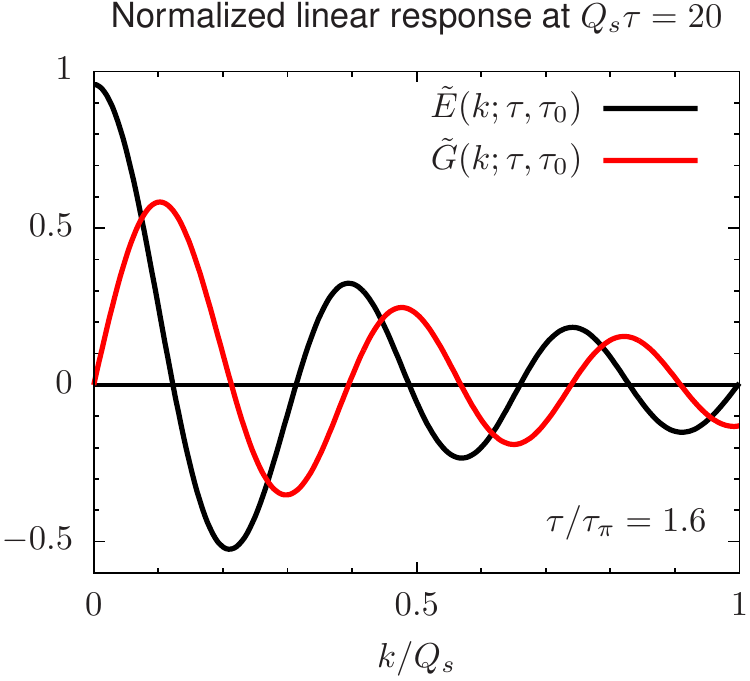}
\caption{Normalized linear response functions in $k$-domain [\Eqs{E} and (\ref{G})] for the initial energy perturbation (a) at $Q_s \tau=10$ and (b) at 
$Q_s \tau=20$ .\label{linear_response}}
\end{figure*}

In order to construct the initial state for hydrodynamics at $\ti$ from a given geometry at $\tau_0$,
the linear response 
of the components of $T^{\mu\nu}$ to the initial perturbation are needed. 
The constitutive relations reduce the number of independent components
of the energy momentum tensor, so it suffices to specify only $\delta T^{00}$ and $\delta T^{0x}$.
\Fig{linear_response}(a) displays the energy  
and momentum  response functions 
($\tilde{E}(k; 
\tau,\tau_0)$ and $\tilde{G}(k; \tau,\tau_0)$  respectively)
to 
an initial energy perturbation $\delta e(\tau_0,k)$ in $\k$-space
\begin{align}
\frac{\delta e(\tau,k)}{ e(\tau)} &\equiv \tilde{E}(k; \tau,\tau_0)\frac{\delta e(\tau_0,k)}{ 
e(\tau_0)},\label{E}\\
\frac{ g^x(\tau,k)}{ e(\tau) }&\equiv-i\tilde{G}(k; 
\tau,\tau_0)\frac{\delta 
e(\tau_0,k)}{ e(\tau_0) }.\label{G}
\end{align}
The results are presented at two suggested initial times $\ti = \{10,20\}/Q_s$.
We will analyze the response functions at asymptotically small $k$ and in coordinate space
in the next two subsections.
Analogous results for the response to an initial momentum perturbation are given in the \app{velpert}. 

\subsection{The kinetic theory response at asymptotically small $k$}
\label{smallk}

The most important contribution to the flow arises from the 
average nuclear geometry, which is smooth on  spatial scales of order $c\ti$.
For this reason the flow due to the average geometry is determined by the $k\rightarrow 0$ limit of the response functions. This section will provide
an analytic understanding of this limit, i.e. the intercept of $\tilde{E}(k; 
\tau,\tau_0)$ and the slope of $\tilde{G}(k; \tau,\tau_0)$ in 
\fig{linear_response}.

First, we will determine how the long wavelength energy perturbations in the transverse plane
change as a function of time.
Returning to the conservation 
equations, \eq{econs} and (\ref{decons}), and setting $k=0$, we have
\begin{align}
&\partial_\tau e(\tau)=-\frac{e(\tau)+
T^{zz}(\tau)}{\tau}, \label{econs2}\\
&\partial_\tau \delta e(\tau)=-\frac{\delta 
e(\tau) +  \delta T^{zz}(\tau)}{\tau}.\label{decons2} 
\end{align}
From these equations the fractional perturbations in the
transverse plane $\delta e/e$ remain constant in time 
in the free streaming limit (where $T^{zz}$ and $\delta T^{zz}$ are
zero), and in the hydrodynamic limit (where $T^{zz}$ and $\delta T^{zz}$ 
are one third  $e$ and $\delta e$). 
Outside
of these limits  $\delta e/e$ is not constant in time.

However, a constant of the motion at $k=0$ can be constructed 
whenever the hydrodynamic gradient expansion is applicable. Indeed, by dimensional analysis, an all order constitutive 
equation at $k=0$  must take the following form
\begin{equation}
T^{zz} = e f(e^{1/4}\tau),
\end{equation}
where $f(x)$ is an order one function and $\delta T^{zz}=\partial_e T^{zz}\delta e$.
Then straightforward steps show
that to all orders in the gradient expansion
\begin{equation}
    \label{detzz}
\lim_{k\rightarrow 0}\frac{\delta e(\tau,k)}{\,3 e- T^{zz}}=\text{const.}
\end{equation} 
For conformally invariant theories with $T^{xx}{=}T^{yy}$ this can be written as
\begin{equation}
    \label{detxx}
    \lim_{k\rightarrow 0}\frac{\delta e(\tau,k)}{e+T^{xx}}=\text{const.}
\end{equation} 

\begin{figure*}
\centering
\includegraphics[width=0.45\linewidth]{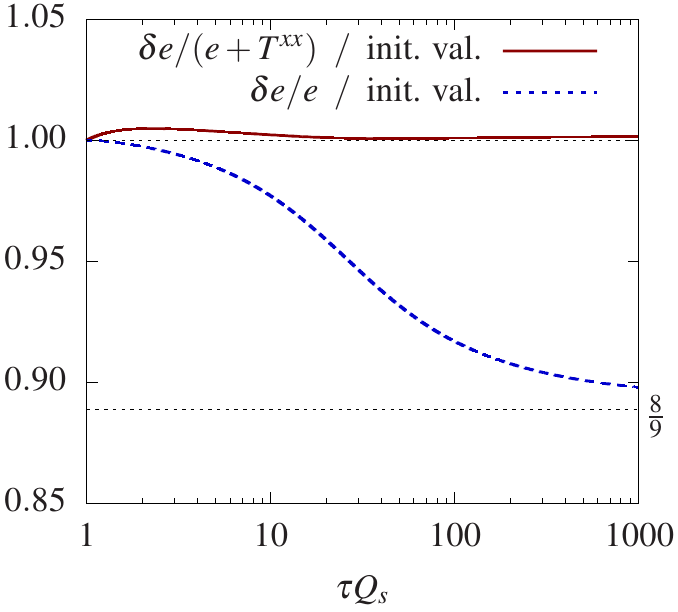}
\hspace{0.03\linewidth}
\includegraphics[width=0.46\linewidth]{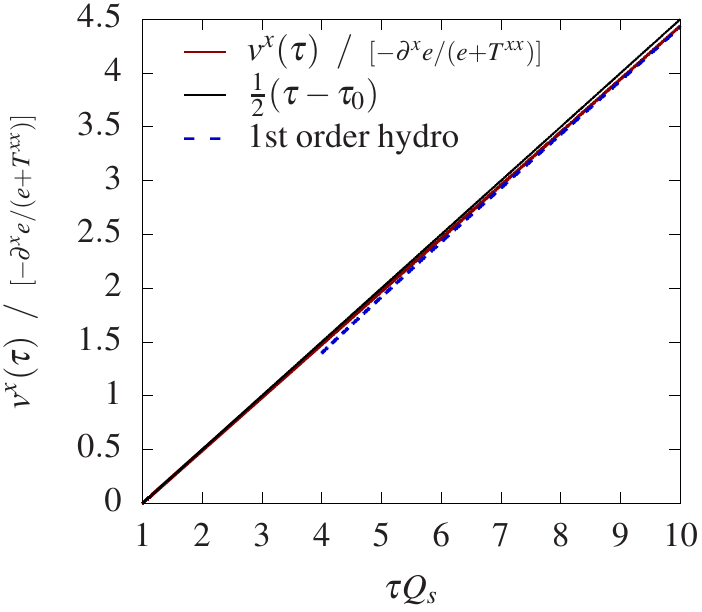}
\caption{ (a) Normalized energy perturbation versus time in the (asymptotically) small $k$ limit. $8/9$ is the 
change in $(e+T^{xx})/e$ between free streaming and ideal hydrodynamic limits (see \eq{eratio}).
(b) The velocity perturbation versus time in the (asymptotically) small $k$ limit scaled
by $-\partial^{x}e /(e+T^{xx})$ (see \eq{vresult2}). The result is compared to 
$\tfrac{1}{2} (\tau-\tau_0)$ (see also ref.~\cite{Vredevoogd:2008id}) and first 
order 
hydrodynamics.}
\label{fig:intde}
\end{figure*}

In \fig{fig:intde}(a) we present the time evolution of $\delta e/e$ and $\delta 
e/(e + T^{xx})$ relative to their initial values. For our initial conditions 
$\delta e /(e + T^{xx})$ remains very nearly constant throughout the entire evolution.
Using this result, the
 change in $\delta e/e$ can be determined by the ratio of 
 $(e+T^{xx})/e$ at the initial and final times, when $T^{xx}$ is approximately $e/2$ and $e/3$ respectively. This reasoning leads to 
 an asymptotic relation between the initial and final energy perturbations
 \st
 \label{eratio}
 \lim_{\tau \rightarrow \infty} \frac{\delta e(\tau)}{e(\tau)} = \frac{8}{9} \frac{\delta e(\tau_0) }{e(\tau_0)},
 \stp
 which is shown in \fig{fig:intde}(a).

Next, we will determine the velocity at (asymptotically) small $k$ as
a function of time from the pre-thermal evolution.
From the conservation equations for perturbations, \Eqs{decons} and 
(\ref{dgcons}),
the  momentum perturbations at small $k$ satisfy
\begin{equation}
    \label{grelation}
\partial_\tau \left(\frac{\tau g^x}{ik} + \frac{1}{2}\delta e 
\tau^2\right)=  
-\frac{\tau}{2} \left(2\delta T^{xx} + \delta T^{zz} -\delta T^{00}\right).
\end{equation}
For conformal theories with $\delta T^{xx}{=}\delta T^{yy}$ the right 
hand side of \eq{grelation} is zero and 
\st
\label{momentumeq}
\frac{\tau g^x}{ik} + \frac{1}{2}\delta e 
\tau^2= \mbox{const}.
\stp
At late times and in coordinate
space this condition reads 
\begin{equation}
\frac{T^{0x}(\tau)}{T^{00}(\tau)}=-\frac{1}{2}\tau\frac{\partial_x 
T^{00}(\tau)}{T^{00}(\tau)} \, ,
\end{equation} 
which was first noted in~\cite{Vredevoogd:2008id}. Here 
we have shown that this relation is a consequence of conformal symmetry (see also \cite{Vredevoogd:2008id}) and the small $k$ limit.

Using \eq{momentumeq} and the definition $g^{x}=(e+T^{xx})v^{x}$, the velocity as a function of time is given by
\begin{equation}
    \label{vresult1}
    \frac{v^x}{ik} = -\frac{\tau}{2} \frac{\delta e}{e + T^{xx} }
\, \left(1-\frac{\delta e(\tau_0)\tau_0^2}{\delta e(\tau)\tau^2}\right).
\end{equation}
Thus, after a brief transient period of order $\tau_{0}$, the velocity 
is directly proportional to time
\st
\label{vresult2}
v^{x} =\frac{\tau}{2} \, \left(\frac{-\partial^x e(\tau, \x)}{e(\tau) + 
T^{xx}(\tau)}\right)\,,  \qquad \frac{-\partial^x e(\tau,\x)}{e(\tau) + 
T^{xx}(\tau)} = {\rm const}.
\stp
In \fig{fig:intde}(b) we compare the growth of the velocity with time
given by \eq{vresult1} with a simple estimate based on \eq{vresult2}.
The simple estimate does a remarkably good job for all times.

\subsection{Response in coordinate space}
\label{response}
To construct the initial conditions for hydrodynamics with 
the correct prethermal evolution, we determine the 
Green functions $E(|\x|;\tau,\tau_0)$ and $G(|\x|;\tau,\tau_0)$ which convert 
the initial  profile of energy perturbations $\delta e(\tau_0,\x)$ to the required 
energy and momentum fluctuations at thermalization time
\begin{subequations}
    \label{kerneleqs}
\begin{align}
\frac{\delta e(\tau,\x)}{e(\tau)} &= 
\int 
d^2\x'
\frac{\delta 
e(\tau_0,\x')}{ e(\tau_0) }
E( 
|\x-\x'|; 
\tau,\tau_0),\\
\frac{ g^i(\tau,\x)}{e(\tau)} &= \int 
d^2\x'
 \frac{\delta 
e(\tau_0,\x')}{ e(\tau_0)}
\frac{(\x-\x')^i}{|\x-\x'|}G( |\x-\x'|; 
\tau,\tau_0).
\end{align}
\end{subequations}
Currently hydrodynamic simulations often smooth the initial conditions
before
starting the hydrodynamic evolution
by convolving the energy density with a Gaussian%
\footnote{See ref.~\cite{Noronha-Hostler:2015coa} for a current
discussion of the observables that are influenced by this arbitrary regulator.}. \Eq{kerneleqs}  by contrast
smooths the initial conditions in a physical way, and 
provides
an attractive alternative to this ad hoc procedure.
\begin{figure*}
\centering
\includegraphics[width=0.47\linewidth]{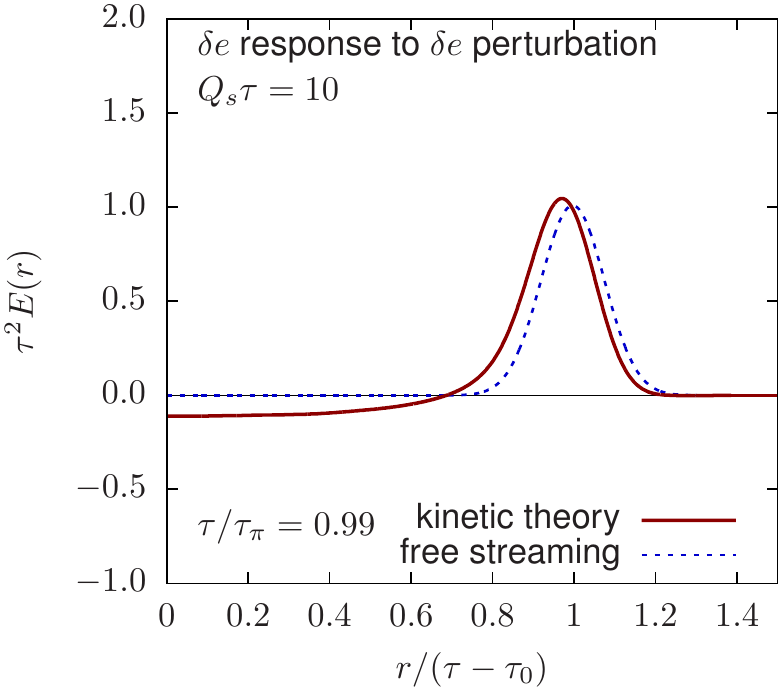}
\includegraphics[width=0.47\linewidth]{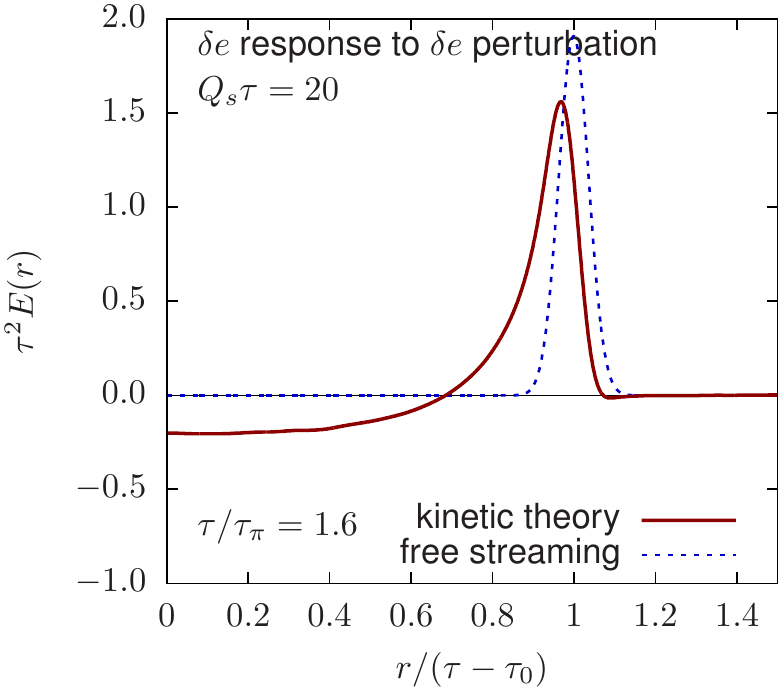}
\includegraphics[width=0.47\linewidth]{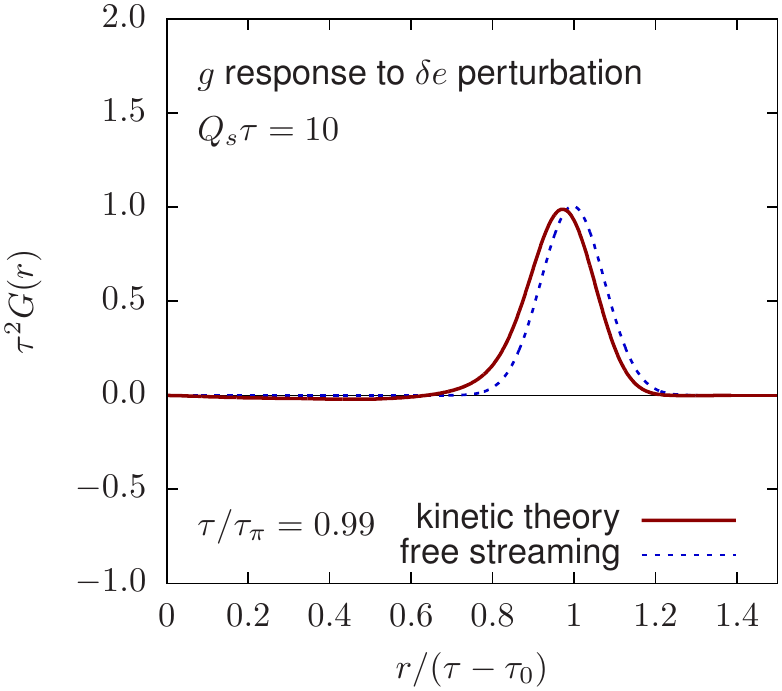}
\includegraphics[width=0.47\linewidth]{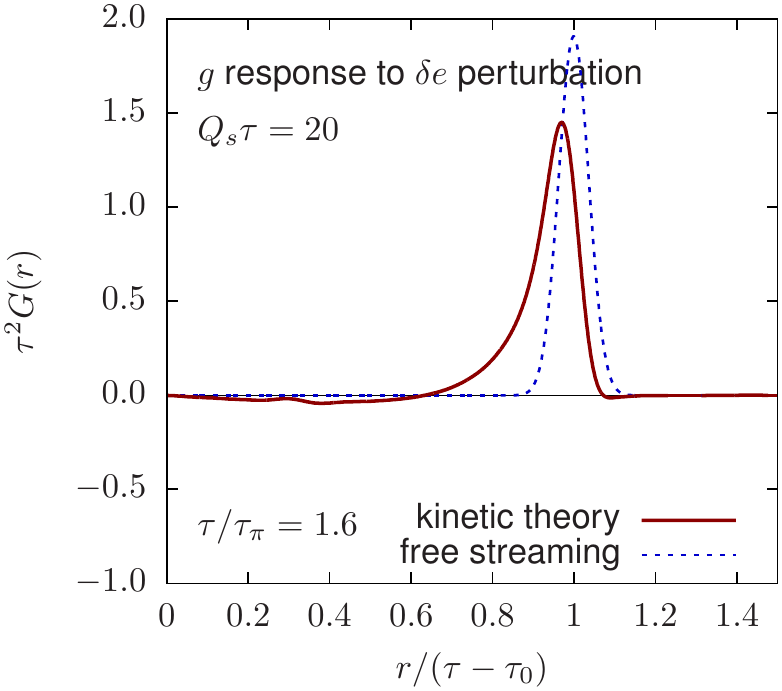}
\caption{(top) Energy and  (bottom) momentum Green functions,     
\Eq{kerneleqs}, for initial \emph{energy perturbation} in coordinate space at 
(left) $Q_s\tau=10$  and (right) $Q_s \tau=20$. 
}
\label{fig:kernelv0}
\end{figure*}

\begin{figure*}
\centering
\includegraphics[width=0.47\linewidth]{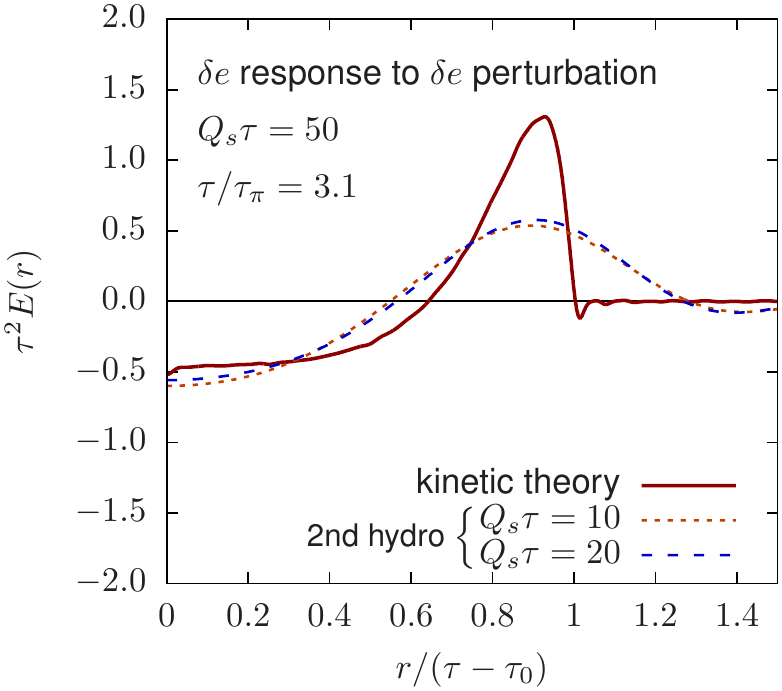}
\includegraphics[width=0.47\linewidth]{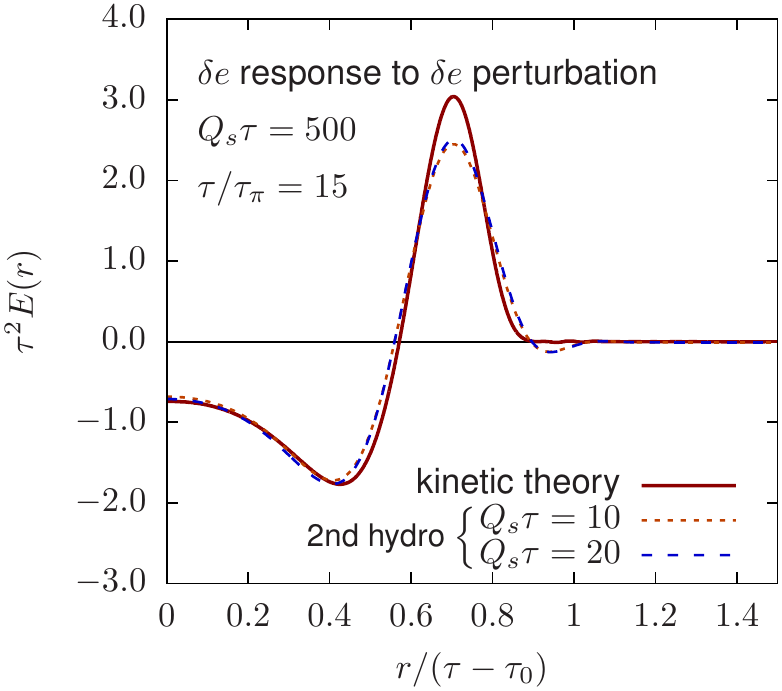}
\caption{Energy Green functions for initial energy perturbations in coordinate space at late times (a) $Q_s \tau=50$ and (b) $Q_s \tau=500$. 
    The results are compared to linearized second order hydrodynamics (\Eqs{eq:conservation} and (\ref{eq:constit})), with the initial
conditions obtained from kinetic theory at $Q_s\tau{=}10$ and $Q_s \tau{=}20$ 
(see \fig{linear_response}). }
\label{fig:kernelv0_late}
\end{figure*}

The EKT is applicable for distance scales  that are larger than the 
 Compton wavelength of the particles $\sim 1/Q_s$. This limits the accuracy of the 
 Green function in spatial domain that can be reached in a computation based on kinetic theory. 
 In order to fold this uncertainty into our result, we regulate our Green function  by convoluting with a Gaussian weight,  $e^{-r^2/2\sigma^2}/(2\pi\sigma^2)$, with $r=|\x|$, and with a 
 width of the order of the initial Compton wavelength $\sigma Q_s = 
 0.7$.
In momentum space this corresponds to suppressing the large $k$ contributions 
by an exponential
envelope $\exp(-  \sigma^2 k^2 / 2)$
\begin{align}
E(|\x|; \tau,\tau_0) &= \int \frac{d^2 \k}{(2\pi)^2} e^{i \k\cdot \x} 
e^{-\sigma^2 k^2/2} E(|\k| ; \tau,\tau_0), \label{ft1} \\
G(|\x|; \tau,\tau_0) &= \int \frac{d^2 \k}{(2\pi)^2}(-i 
\hat{k}\cdot\hat{x}) e^{i 
\k\cdot \x} 
e^{-\sigma^2k^2/2} G(|\k| ; \tau,\tau_0) \label{ft2}.
\end{align}
The regulated Green functions  are shown in \fig{fig:kernelv0} at the  
initialization times $\ti Q_s=\{10,20\}$
(for details of the Fourier transform see \app{kernelFT}.) At $\tau Q_s=10$ the 
system has spent a significant proportion of the total evolution time with small
longitudinal pressure
$T^{zz}\approx 0$, and therefore the resulting response is similar to the free 
streaming prediction (see \app{kernelFT}). However, the Green function  in 
\fig{fig:kernelv0}(a) 
is peaked for $r{<}c|\tau-\tau_0|$,  suggesting  a slight deflection 
from the free streaming trajectory.
Additionally,  
the energy perturbation is negative at small $r$,  which is indicative 
of a nascent approach to hydrodynamics. At later times, such as $Q_s\tau=20$ in 
\fig{fig:kernelv0}(b), 
these differences become more pronounced. 
Similar features are visible in the momentum response to an initial energy 
perturbation shown \fig{fig:kernelv0}(c) and (d).
Finally, in \fig{fig:kernelv0_late} we 
show 
the Green functions at later times $Q_s\tau=50$ and $Q_s\tau=500$, 
and compare to linearized  second order hydrodynamics (\Eqs{eq:conservation} and (\ref{eq:constit})) with initial conditions
taken from the $Q_s\tau=\{10,20\}$ results.  Between $Q_s\tau{=}20$ and $Q_s\tau{=}50$, the hydrodynamics overdamps the high $k$ modes  (see also \cite{Romatschke:2015gic}), and the response
is broader than the predictions of kinetic theory. However,
these Green functions will be convolved with the initial conditions,  
and thus the resulting hydrodynamic initial state is mostly sensitive to the first moments of
these kernels. The moments of the EKT and hydro kernels are
determined by the small $k$ behaviour of the response functions, which
agree to a few percent (not shown). At later times $Q_s\tau=500$, the
response is largely determine by Fourier modes in the hydrodynamic 
regime $k \lesssim 0.1\,Q_s$, and the EKT and hydro kernels are visually similar.

To summarize, the hydrodynamic evolution sets in early at rather large
anisotropies, and the hydrodynamic constitutive equations are approximately satisfied as soon as the
$T^{zz}$ starts to significantly deviate from the free streaming expectation,
$T^{zz}\approx 0$. For this reason 
the time interval when the evolution is not described by 
free streaming or hydrodynamics is comparatively brief (see 
\fig{background}(a)),
and as hydrodynamics becomes marginally applicable at $Q_s\ti \sim 10$, the Green function closely resembles the free streaming result.
Therefore, an approach where the evolution is described by free streaming until $\ti$ 
seems well motivated \cite{Broniowski:2008qk,Liu:2015nwa}, provided that the correct value of $\ti$ is used. However, such  
an ad hoc approach does not account for some of the qualitative details of the Green function, such as the depletion of the energy density in the interior region.

\section{Discussion}
\label{discuss}
In this paper, we have provided a bridge between the far-from-equilibrium
initial conditions of heavy-ion collisions and hydrodynamized plasma. Our main
result is the coordinate space Green functions (see \Eq{kerneleqs} and
\fig{fig:kernelv0}), which can be used to filter the pre-equilibrium
energy density to find the full energy-momentum tensor for hydrodynamics at the
initialization time.  The procedure can be implemented in complete hydrodynamic
simulations, removing one source of uncertainty.  Perhaps more importantly, the
approximations in the EKT are compatible with the IP-Glasma setup, and thus the
whole evolution from saturated nuclei to hydrodynamics can be comprehensively
modelled within a perturbatively controlled framework.

We provide the coordinate space Green functions at two different suggested initialization times, $Q_s \ti=\{10,20\}$. 
At the earlier initialization time, $Q_s\ti=10$,
there are  significant (though bearable) corrections 
to the constitutive relations due to non-hydrodynamic modes (see 
\fig{background}(a) and \fig{figdTxx}). By $Q_s\ti=20$
the constitutive relations at small $k$ are well satisfied, and the subsequent
evolution is reasonably
captured by  second order hydrodynamics%
\footnote{When examining \fig{fig:kernelv0_late}, one must remember that the 
full Green functions
will be convolved with the initial conditions, and thus the response
of the system is mostly sensitive to the first moments of the  
kernels in \fig{fig:kernelv0_late}.
The first EKT moments (i.e. the small $k$ behavior of the response) agree with the hydrodynamics to the percent level. }
(see \fig{fig:kernelv0_late}).
The approximate overlap of the two 2nd order viscous lines in 
\fig{fig:kernelv0_late}, which correspond to initializing the hydro at $\ti 
Q_s= \{10,20\}$,
demonstrates that the subsequent hydrodynamical evolution is indeed rather insensitive to the initialization time. 
In \sect{Green} we examined the (asymptotically) small $k$ 
limit of the Green functions, and confirmed (and clarified)
a preflow estimate by Vredevoogd and Pratt~\cite{Vredevoogd:2008id} (see 
\fig{fig:intde}).

The hydrodynamics that the EKT follows is characterized by the weak coupling
value of $\eta/s \approx 0.62$~\cite{Arnold:2003zc},  which is
significantly higher than the AdS/CFT result $\eta/s\simeq 0.08$~\cite{Policastro:2001yc},  and
current phenomenological estimates, which assume that
$\eta/s$ is independent of the temperature.
Recent analyses have relaxed the temperature independence of $\eta/s$, and
shown that 
the value of $\eta/s$ at higher temperatures $T \sim 3 T_c$ is poorly
constrained by data~\cite{Niemi:2015qia}. Since it is the high temperature regime that is most relevant for the transition to hydrodynamics,  we believe
that the current kinetic theory results for the initial stages can be
consistent with hydro phenomenology, provided $\eta/s$ decreases towards the
strong coupling result as the system cools towards $T_c$.

Nevertheless, to apply our results to a hydro simulation with  lower viscosity than
the perturbative expectation, we note that the two  initialization
times $\ti Q_s = \{10,20\}$ correspond in units of the hydro parameters  to $\ti = \{.99,1.6\} \tau_\pi$.
The scaled times $\{0.99,1.6\}$ can be used 
to initialize simulations when the transport coefficients differ.
Such an approach is supported by the reasonably good
scaling properties of the hydrodynamization times and prethermal evolution as
function of the coupling constant when expressed in terms of the hydrodynamic
variables~\cite{Kurkela:2015qoa,Keegan:2015avk}. 

Although our EKT description can be further improved by inclusion of fermionic
degrees of freedom and by improving the connection to the early classical
evolution, we believe that it already provides a physically sound picture of
the approach to hydrodynamics and can be used to initialize all components of
the energy-momentum tensor for subsequent hydrodynamic evolution. This
eliminates a source of uncertainty in current simulations, and provides a satisfyingly complete
description of the early time evolution in heavy ion collisions.

\acknowledgments
The authors thank Fran\c cois Gelis, Keijo Kajantie, Tuomas Lappi, Risto Paatelainen, Jean-Fran\c cois Paquet, and Urs Wiedemann for useful discussions.
The simulations were performed using computer resources of UiS and  
high-performance LIred computing system at the Institute for Advanced 
Computational Science at Stony Brook University. 
A.M. and D.T. work was supported in part by the U.S. Department of Energy under 
Contracts No. DE\nobreakdash-FG\nobreakdash-88ER40388. Finally, A.M. and D.T 
would like to thank CERN Theoretical Physics Department for the hospitality 
during the short-term visit.

\appendix

\section{Collision kernel}
\label{colker}
In this appendix we provide additional details on the collision
kernels used in  \eq{boltz}. The collision kernel for the uniform 
background 
contains terms arising  from 
elastic $2 \leftrightarrow 2$ scatterings and inelastic $1\leftrightarrow 2$ 
collinear splittings
\begin{equation}
\mathcal{C}[f] = \mathcal{C}_{2 \leftrightarrow 2}[f] + \mathcal{C}_{1 \leftrightarrow 2}[f].
\end{equation}
The two collision terms read\,~\cite{Arnold:2002zm, 
Kurkela:2015qoa,Keegan:2015avk}
\begin{align}
\label{2to2}
 \mathcal{C}_{2\leftrightarrow 2}[f](\p)&= \frac{1}{4|\p| \nu_g}\int \frac{d^3 k}{2k (2\pi)^3}\frac{d^3 p'}{2p' (2\pi)^3}\frac{d^3 k'}{2k' (2\pi)^3}|\mathcal{M}(\p,\k;\p',\k')|^2 (2\pi)^4 \delta^{(4)}(P+K-P'-K') \nonumber \\
 &\times \big\{ f_\p f_\k[1+ f_{\p'}][1+ f_{\k'}]-f_{\p'}f_{\k'}[1+ f_{\p}][1+ 
 f_{\k}] \big\}
\end{align}
and
\begin{align}
\label{1to2}
\mathcal{C}^{1\leftrightarrow 2}[f](\p) &= \frac{(2\pi)^3}{2|\p|^2 
\nu_g}\int_0^{\infty} dp' dk' \, \delta(|\p|-p' 
-k')\gamma(\p;p'\hat{\p},k'\hat{\p})\nonumber 
\times
\big\{f_\p[1+ f_{p' \hat{\p}}][1+ f_{k' \hat{\p}}] - f_{p' \hat{\p}}f_{k' \hat{\p}}[1+ f_{\p}]\Big\}\nonumber\\
&+ \frac{(2\pi)^3}{|\p|^2 \nu_g}\int_0^{\infty} dp' dk \, \delta(|\p|+k 
-p')\gamma(p' \hat{\p};\p,k \hat{\p})
\times
\big\{ f_{\p}f_{k \hat{\p}}[1+ f_{p' \hat{\p}}] -f_{p' \hat{\p}}[1+ f_\p][1+ f_{k\hat{\p}}]\Big\},
\end{align}
where $\hat{\p}$ is the unit vector parallel to $\p$, and capital letters 
denote null 4-vectors, i.e. $P^0 \equiv |\p|$. The effective elastic 
$|\mathcal{M}|^2$ and inelastic $\gamma$ scattering matrix elements contain 
non-trivial structures arising from the soft and collinear divergences, which  
are dynamically regulated by the in-medium physics. 

For the most 
of kinematics the effective elastic scattering element is given 
by\footnote{\label{fnote}  Equations \Eq{c2to2} and 
\Eq{spliting} have some minor typos corrected compared to 
refs.~\cite{Kurkela:2015qoa,Keegan:2015avk}.}
\begin{align}
|\mathcal{M}|^2 = 2 \lambda^2 \nu_g \left( 9 + \frac{(s-t)^2}{u^2}+ 
\frac{(u-s)^2}{t^2}+ \frac{(t-u)^2}{s^2}\right).\label{c2to2}
\end{align}
For a soft gluon exchange  with the momentum 
transfer $q = |\p' - \p|$  in $t$-channel (or $q = |\p' - \k|$ in $u$-channel)
the collision 
matrix is proportional to 
$\propto 1/(q^2)^2$, and thus suffers from a soft Coulomb divergence.  It is 
regulated by replacing 
\begin{equation}
q^2 t \rightarrow (q^2 + 2 \xi_0^2 m^2)t,
\end{equation}
in the denominators of divergent terms (similarly for the $u$-channel). Here 
$m^2$ is the thermal asymptotic 
mass of the gluon defined as
\begin{equation}
m^2 = 2 \lambda \int \frac{d^3 p}{(2\pi)^3}\frac{f_\p}{|\p|}.
\end{equation}
The coefficient $\xi_0= e^{5/6}/\sqrt{8}$ is fixed so that the matrix element 
reproduces the 
drag and momentum diffusion properties of soft scattering at leading order for isotropic distributions $f_{\p}$ \cite{York:2014wja}.  

The effective splitting kernel reads 
\begin{align}
\gamma(p {\bf \hat p}; p' {\bf \hat p},k' {\bf \hat p}) = \frac{p^4 + 
p'^4+k'^4}{p^3 p'^3 k'^3 }\frac{\nu_g \lambda}{8 (2\pi)^4}\int \frac{d^2 
h}{(2\pi)^2} 2 {\bf h} \cdot {\rm Re} {\bf F},\label{c1to2}
\end{align} 
where the equation for ${\bf F}$ accounts for splitting due to multiple 
scatterings with transverse momentum exchange ${\bf} q$, and momentum 
non-collinearity
\begin{align}
2 {\bf h} = &  i \delta E({\bf h} ){\bf F}({\bf h})   +\frac{\lambda 
T_*}{2}\int \frac{d^2  q_\perp}{(2\pi)^2} \mathcal{A}({\bf 
q}_\perp)\label{spliting}  \\ & 
\times \Big[ 3{\bf F({\bf h})}-{\bf F}({\bf h}-p' {\bf q}_\perp)-{\bf F}({\bf 
h}-k' {\bf q}_\perp)-{\bf F}({{\bf h}+p {\bf q}_\perp})\Big]\nonumber.
\end{align}
with
$
T_* = \frac{\lambda}{m^2}\int \frac{d^3 p}{(2\pi)^3}f_\p(1+f_\p) ,
$
and $\delta E = m^2/2p'+m^2/2k'-m^2/2p + {\bf h}^2/2p k' p'$. In 
the isotropic screening approximation
\begin{equation}
\mathcal{A}({\bf q}_\perp) = \left( \frac{1}{{\bf q}_\perp^2} - \frac{1}{{\bf 
q}_\perp^2 + 2m^2}\right).
\end{equation}
Both $m^2$ and $T_*$ are self-consistently evaluated at 
each time step.

The linearized collision kernels are obtained trivially  by replacing $f 
\rightarrow \bar f + \delta f$ in the integrands of \ref{2to2} and  
\ref{1to2}  and 
linearizing in $\delta f$. In addition one has to take into account the 
linear variation of the thermal mass $\delta m^2 $ and the effective 
temperature 
$\delta T_*$ in the scattering matrix elements \Eqs{c2to2} and (\ref{c1to2})
\begin{align}
\delta m^2 &= 2 \lambda \int \frac{d^3 p}{(2\pi)^3} \frac{\delta f_\p}{|\p|},\\
\delta T_*  &= \frac{\lambda}{m^2} \int \frac{d^3 p}{(2\pi)^3}  \delta f_\p 
(1+ 2 f_\p) -\frac{\delta m^2}{m^2} T_*.
\end{align}
where $m^2$ and $T_* $ are evaluated from the unperturbed  background 
distribution.

\section{Fourier transform of Green functions}
\label{kernelFT}
Here we provide details of performing Fourier transforms in \Eqs{ft1} and (\ref{ft2}) to obtain spatial Green  functions shown in 
\figs{fig:kernelv0} and \ref{fig:kernelv0_late}. The two dimensional Fourier 
transforms can be straightforwardly reduced to 
one dimensional
Hankel transforms
\begin{align}
E( |\x|; \tau,\tau_0) &= \int_0^\infty \frac{dk}{2\pi} k \tilde{E}(k; 
\tau,\tau_0) 
e^{-\sigma^2 k^2/2}
J_0(k 
|\x|),\label{H1}\\
G(|\x|; \tau,\tau_0) &= \int_0^\infty \frac{dk}{2\pi} k  \tilde{G}(k; 
\tau,\tau_0) 
e^{-\sigma^2 k^2/2}\label{H2}
J_1(k |\x|).
\end{align}
Integrals in \Eqs{H1} and (\ref{H2})  were done numerically by using cubic 
interpolation for $\tilde{E}$ 
and $\tilde{G}$ within the available range of wavenumbers $k\in [0,4]Q_s$ 
To avoid the oscillatory behaviour due to a sharp $k$ cut-off at $k=4Q_s$, we 
extrapolated the Green functions until the Gaussian envelope
$e^{-\sigma^2 k^2/2}$ 
smoothly cuts off the integral. 
 For extrapolation at large $k$ we used functional forms motivated by free streaming results: $C_0 J_{0}(v_0|k|(\tau-\tau_0))$ and $C_1 J_{1}(v_1|k|(\tau-\tau_0))$, where coefficients 
$C_i$ and $v_i$ were fitted to match the oscillatory  behaviour of Green functions at the largest available $k$.
For $Q_s\tau=\{10,20,50\}$ we used $Q_s \sigma = 0.7$ for the envelope 
corresponding roughly to the smallest scales the EKT can resolve. 
 For $Q_s \tau=500$, perturbations with large wavenumbers were sufficiently 
 suppressed by EKT evolution that no extrapolation was 
 necessary.

For early times and large values of $k$ the collision terms in the Boltzmann 
equation \eq{boltz} can be neglected and the system is freely streaming. For 
particle distributions that are highly anisotropic in $z$ direction 
($\mathcal{P}_L\ll \mathcal{P}_T$), but isotropic in $xy$-plane, energy 
perturbations are propagating in circular wavefronts at the velocity $v$ of constituent particles  (for massless gluons $v=c$). In such free streaming evolution energy perturbations at time $\tau$ and position $\x$ are equal to the average of energy perturbations at $\tau_0$ on a circle
$|\x-\x'|=c|\tau-\tau_0|$\cite{Liu:2015nwa}. Thus, free streaming Green functions in coordinate space are
\begin{equation}
E( |\x|; \tau,\tau_0)=G( |\x|; \tau,\tau_0) = \frac{1}{2\pi 
|\x|}\delta(|\tau-\tau_0|-|\x|).
\end{equation}
Here we also quote the free streaming result for the response to initial momentum perturbations defined in Appendix \ref{velpert}
\begin{align}
E(|\x|,\tau,\tau_0) &=\frac{2}{2\pi |\x|} \delta(|\tau-\tau_0|-|\x|),\label{freev1}\\
G(|\x|,\tau,\tau_0) &=\frac{2}{2\pi |\x|} \delta(|\tau-\tau_0|-|\x|)-\frac{2}{2\pi}
\frac{1}{|\x|}\theta(|\tau-\tau_0|-|\x|).\label{freev2}
\end{align}
Free streaming Green functions shown in \fig{fig:kernelv0} and  
\fig{fig:kernelv0_g} were also folded in with a Gaussian regulator as 
discussed  above.

\section{Initial velocity perturbations}
\label{velpert}
\begin{figure*}
\centering
\includegraphics[width=0.47\linewidth]{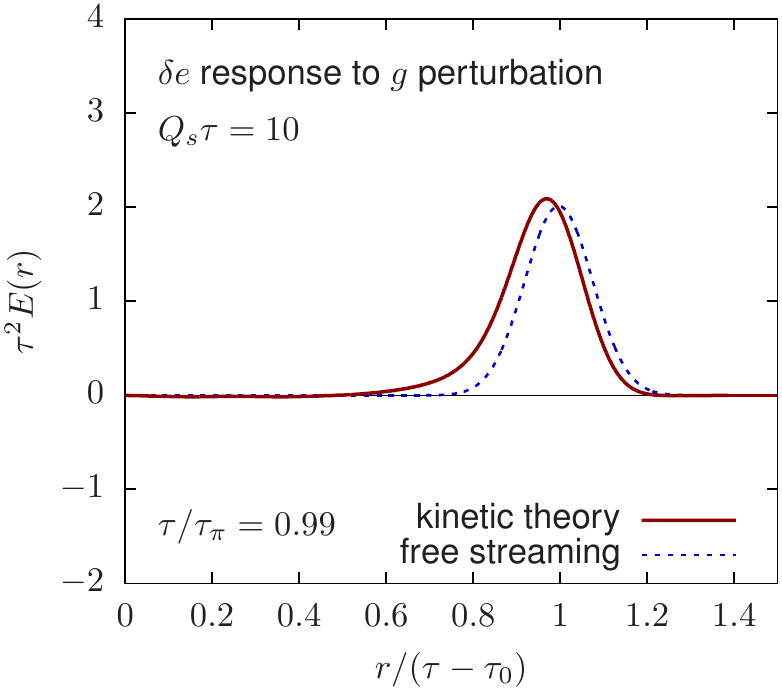}
\includegraphics[width=0.47\linewidth]{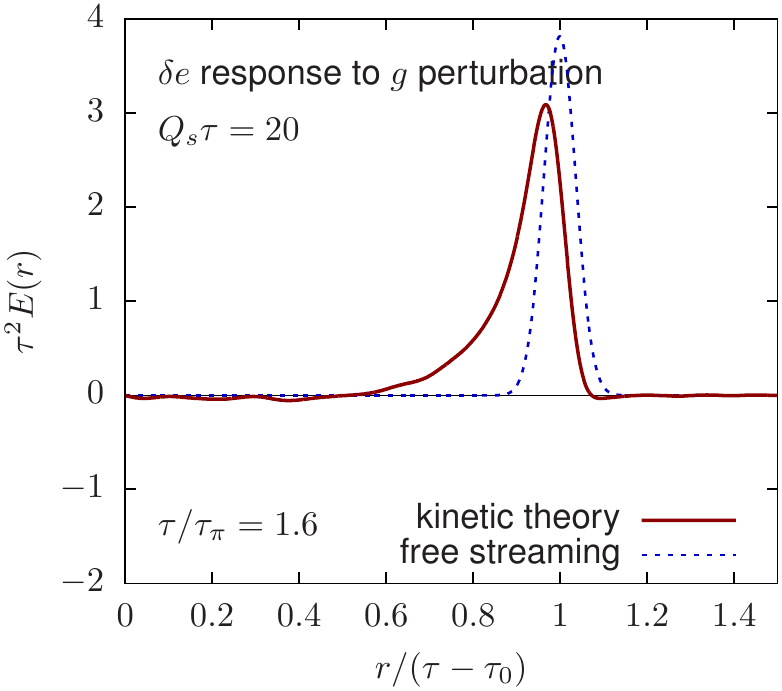}\\
\includegraphics[width=0.47\linewidth]{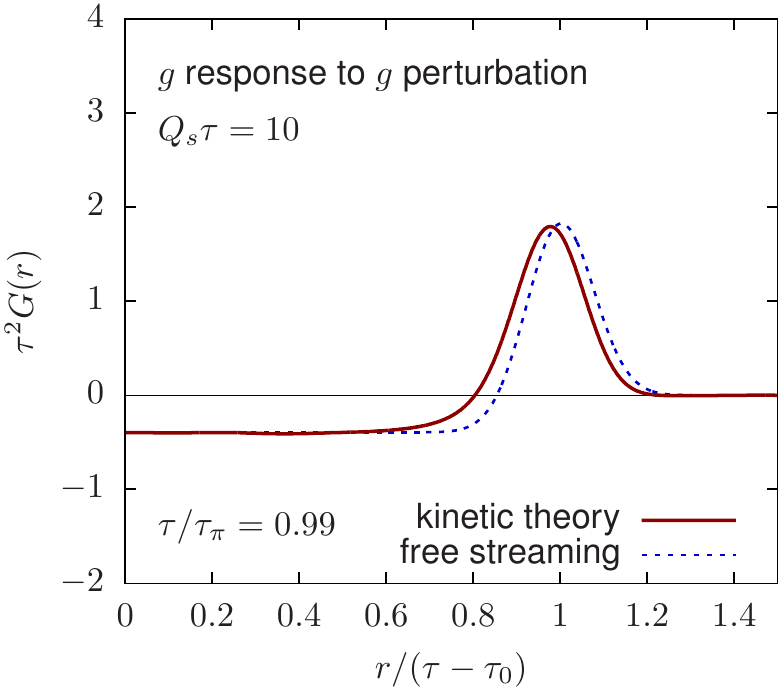}
\includegraphics[width=0.47\linewidth]{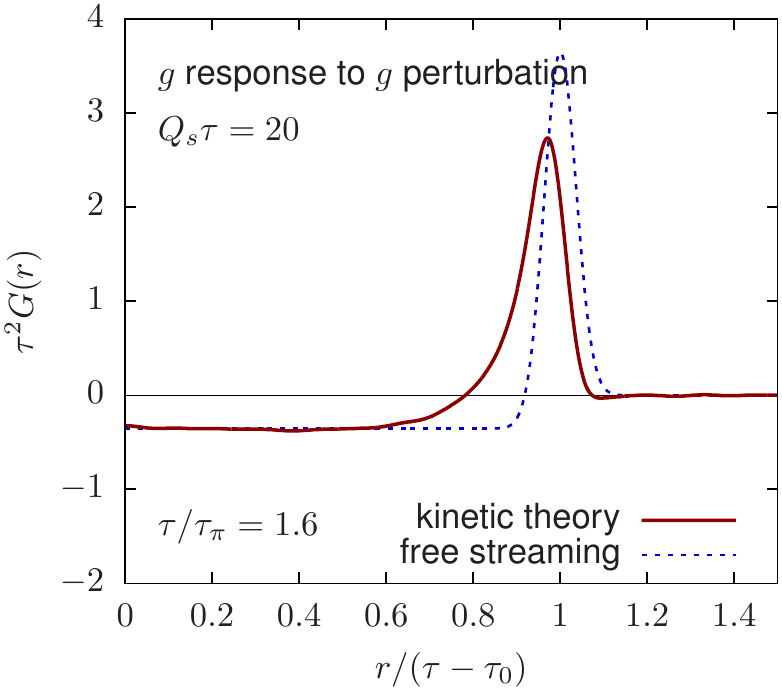}
\caption{(top) Energy and  (bottom) momentum Green functions,     
\Eq{kerneleqs_g}, for initial \emph{momentum perturbation} in coordinate space 
at (left) $Q_s\tau=10$  and (right) $Q_s \tau=20$. 
}
\label{fig:kernelv0_g}
\end{figure*}

In this appendix we summarize results for the EKT response to initial momentum 
perturbation, \Eq{initialdf2}. Mirroring the discussion in \sect{Green}, we 
define Green functions for initial momentum perturbations in $k$-space as 
follows
\begin{align}
\frac{\delta e(\tau,k)}{ e(\tau)} &= -i \tilde{E}(k; 
\tau,\tau_0)\frac{ 
g^x(\tau_0,k)}{ 
e(\tau_0)},\\
\frac{  g^x(\tau,k)}{ e(\tau) }&= \tilde{G}(k; 
\tau,\tau_0)\frac{ 
g^x(\tau_0,k)}{ e(\tau_0) }.
\end{align}
By performing Fourier transform and regularization (see \app{kernelFT} for details) we obtain Green functions in coordinate space, which can be used to propagate initial momentum fluctuations to energy and momentum perturbations at a given thermalization time
\begin{subequations}
    \label{kerneleqs_g}
\begin{align}
\frac{\delta e(\tau,\x)}{e(\tau)}  
 &= 
\int 
d^2\x'
\frac{(\x-\x')^i}{|\x-\x'|}\frac{ 
g^i(\tau_0,\x')}{ e(\tau_0) }
E( 
|\x-\x'|; 
\tau,\tau_0),\label{gt1}\\
\frac{g^i(\tau,\x)}{e(\tau)} 
&= \int 
d^2\x'
 \frac{g^i(\tau_0,\x')}{ e(\tau_0)}
G( |\x-\x'|; 
\tau,\tau_0).\label{gt2}
\end{align}
\end{subequations}
The regulated Green functions for initial momentum perturbations are shown in 
\fig{fig:kernelv0_g}.  As a comparison we also include Green functions for free 
streaming 
evolution, \Eqs{freev1} and (\ref{freev2}). 

Energy and momentum Green functions shown in \fig{fig:kernelv0} and 
\fig{fig:kernelv0_g} make a complete set of response functions necessary to 
propagate arbitrary initial energy and momentum perturbations from the 
far-from-equilibrium initial conditions to times $Q_s\tau=\{10,20\}$ when 
hydrodynamics  becomes applicable. Crucially, at these times one can use 
hydrodynamic constitutive equations \Eq{eq:constit} to initialize all 
components of the energy momentum tensor.

\bibliographystyle{JHEP}
\bibliography{preflow}

\end{document}